\documentclass{llncs}
\usepackage{amsmath,amssymb,amsfonts}
\usepackage{tikz} 
\usepackage{color}
\usepackage{xcolor}
\usepackage{xspace}
\usepackage{microtype}
\usepackage{todonotes}
\usepackage{colonequals}
\usepackage{algorithm}
\usepackage{newalg}
\usepackage{multirow}
\usepackage{multicol}
\newcounter{oldlinenumber}
 \usepackage{booktabs}
\usepackage{float}
\usepackage{subfigure}
\usepackage{url}
\usepackage{appendix}
\usepackage{graphicx}
\usepackage{listings}
\usepackage[pdfpagelabels=true]{hyperref}

\usetikzlibrary{arrows,decorations.pathmorphing,positioning,fit,trees,shapes,shadows,automata} 

\newif\ifcomment\commenttrue %funktioniert nicht immer, wenns genestet ist
\newif\iflong\longfalse
\newif\ifshort\shorttrue
\newif\ifaewl\aewltrue %explanation of Algorithm 2 with line specification

\newcommand{\pol}{g}
\newcommand{\altpol}{h}
\newcommand{\bases}[1]{\ensuremath{\mathrm{bases}(#1)}}
\newcommand{\linex}[1]{\ifaewl~(line~\ref{#1})\fi}
\newcommand{\linexx}[2]{\ifaewl~(line~\ref{#1}-\ref{#2})\fi}

\definecolor{lgrey}{rgb}{0.8,0.8,0.8}
\definecolor{grey}{rgb}{0.5,0.5,0.5}

\definecolor{lightblue}{rgb}{0.8,0.8,1.0}
\definecolor{lightred}{rgb}{1.0,0.8,0.8}
\definecolor{lightgreen}{rgb}{0.8,1.0,0.8}
\definecolor{angrygreen}{cmyk}{0.279,0,0.91,0.08}
\definecolor{lightred}{rgb}{1.0,0.8,0.8}
\definecolor{pink}{rgb}{1.0,0.1,1.0}

\newcommand{\sinit}{\ensuremath{s_{\mathrm{I}}}}
\newcommand{\Sinit}{\ensuremath{S_{\mathrm{I}}}}
\newcommand{\starget}{\ensuremath{t}}
\newcommand{\Inp}{\ensuremath{\mathrm{Inp}}}
\newcommand{\Out}{\ensuremath{\mathrm{Out}}}

\newcommand{\abs}{\ensuremath{_\mathrm{abs}}}
\newcommand{\kabs}{\ensuremath{_{K\mapsto \mathrm{abs}}}}

\newcommand{\paths}{\ensuremath{\mathrm{Paths}}}
\newcommand{\pathsfin}{\ensuremath{\paths}}
\newcommand{\pfin}{\ensuremath{\pr}}

\newcommand{\pr}{\ensuremath{\mathrm{Pr}}}

\newcommand{\Rloop}{\ensuremath{R_{\mathrm{loop}}}}
\newcommand{\Rout}{\ensuremath{R_{\mathrm{out}}}}
\newcommand{\ratfunc}{\ensuremath{\mathcal{F}_V}}
\newcommand{\evalratfunc}[2]{\ensuremath{#1[u]}}
\newcommand{\R}{\ensuremath{\mathbb{R}}}
\newcommand{\N}{\ensuremath{\mathbb{N}}}

\newcommand{\Z}{\ensuremath{\mathbb{Z}}}

\newcommand{\DTMCD}[1][]{\ensuremath{\mathcal{D}{#1}}\xspace}

\newcommand{\DTMCdef}[1][]{\ensuremath{\DTMCD{#1}=(S{#1}, I{#1}, P{#1})}\xspace}
\newcommand{\DTMCudef}[1][]{\ensuremath{\DTMCD{#1}=(S_u{#1}, I_u{#1}, P_u{#1})}\xspace}
\newcommand{\DTMCMu}[1][]{\ensuremath{\mathcal{M}_u{#1}}\xspace}
\newcommand{\PDTMCM}[1][]{\ensuremath{\mathcal{M}{#1}}\xspace}

\newcommand{\PDTMCdef}[1][]{\ensuremath{\PDTMCM{#1}=(S{#1},  V{#1}, I{#1}, P{#1})}\xspace}

\newcommand{\graph}{\ensuremath{\mathcal{G}}}

\newcommand{\series}[2]{\ensuremath{#1_1,\ldots, #1_{#2}}}
\newcommand{\seriesPath}[2]{\ensuremath{#1_1\ldots #1_{#2}}}

\newcommand{\ie}{i.\,e., }
\newcommand{\eg}{e.\,g., }

\newcommand{\ginac}{\texttt{GiNaC}\xspace}

\newcommand{\fmult}{\ensuremath{\mathbin{\cdot_{\!\scriptscriptstyle\mathcal{F}}}}}
\newcommand{\fcm}{\ensuremath{\mathbin{\cup_{\scriptscriptstyle\mathcal{F}}}}} %common multiple
\newcommand{\fdiv}{\ensuremath{\mathbin{\colon_{\!\!\!\scriptscriptstyle\mathcal{F}}}}}
\newcommand{\fcd}{\ensuremath{\mathbin{\cap_{\scriptscriptstyle\mathcal{F}}}}} % common divisor
\newcommand{\fplus}{\ensuremath{\mathbin{+_{\!\scriptscriptstyle\mathcal{F}}}}}

\newcommand{\factor}[1]{\ensuremath{\mathcal{F}_{#1}}}
\newcommand{\product}[1]{\ensuremath{\mathcal{P}(#1)}}
\newcommand{\cgcd}{\ensuremath\mathrm{gcd}}

\newcommand{\nullpoly}{0}
\newcommand{\nullnum}{0}
\newcommand{\homepage}{\url{http://goo.gl/nS378q}}

\newcommand{\TO}{{TO}}
\newcommand{\MO}{{MO}}

\makeatletter
\DeclareRobustCommand{\cpp}
{\valign{\vfil\hbox{##}\vfil\cr
   \textsf{C\kern-.1em}\cr
   $\hbox{\fontsize{\sf@size}{0}\textbf{+\kern-0.05em+}}$\cr}\xspace%
}
\makeatother

\newenvironment{mytheorem}[2][]{%
  \par\vspace{0.5\baselineskip}\noindent\textbf{Theorem~#2 \ifthenelse{\equal{#1}{}}{\relax}{(#1) }}\itshape}{%
  \par\vspace{0.25\baselineskip}
}

\begin{document}
  \title{Accelerating\\ Parametric Probabilistic Verification %
\thanks{This work was partly supported by the German Research Council (DFG) as
part of the Research Training Group
AlgoSyn (1298) and the Transregional Collaborative Research Center AVACS (SFB/TR~14), the EU FP7-project MoVeS, the FP7-IRSES project MEALS and by the Excellence Initiative of the German federal and state government.}
}

\author{Nils Jansen\inst{1}\and 
        Florian Corzilius\inst{1}\and 
        Matthias Volk\inst{1}\and 
        Ralf Wimmer\inst{2}\and\\ 
        Erika \'Abrah\'am\inst{1}\and 
        Joost-Pieter Katoen\inst{1}\and        
        Bernd Becker\inst{2}
}

\institute{RWTH Aachen University, Germany \\
           \email{\{nils.jansen | corzilius | volk | abraham |
katoen\}@cs.rwth-aachen.de}
           \and 
           Albert-Ludwigs-University Freiburg, Germany \\
           \email{\{wimmer | becker\}@informatik.uni-freiburg.de}
}

\maketitle

  \begin{abstract}
We present a novel method for computing reachability probabilities of parametric
discrete-time Markov chains whose transition probabilities are fractions of
polynomials over a set of parameters. Our algorithm is based on two key
ingredients: a graph decomposition into strongly connected subgraphs combined
with a novel factorization strategy for polynomials. Experimental evaluations
show that these approaches can lead to a speed-up of up to several
orders of magnitude in comparison to existing approaches.
\end{abstract}

  \section{Introduction}
\label{sec:introduction}

\emph{Discrete-time Markov chains} (\emph{DTMCs}) are a widely used modeling formalism for 
systems exhibiting probabilistic behavior. Their applicability ranges from 
distributed computing to security and systems biology. Efficient algorithms 
exist to compute measures like: ``What is the probability that our communication 
protocol terminates successfully if messages are lost with probability 0.05?". 
However, often actual system parameters like costs, faultiness, reliability and 
so on are not given explicitly. For the design of systems incorporating random 
behavior, this might even not be possible at an early design stage. In 
model-based performance analysis, the research field of 
\emph{fitting}~\cite{rosenblum_fitting}, where---intuitively---probability 
distributions are generated from experimental measurements, mirrors the 
difficulties in obtaining such concrete values.

This calls for treating probabilities as parameters and motivates to
consider \emph{parametric} DTMCs, PDTMCs for short, where transition
probabilities are (rational) functions in terms of the system's
parameters. Using these functions one can, \eg find appropriate values
of the parameters such that certain properties are satisfied or
analyze the sensitivity of reachability probabilities to small changes
in the parameters. Computing reachability probabilities for 
DTMCs is typically done by solving a linear equation system. This is
not feasible for PDTMCs, since the resulting equation system is
non-linear. Instead, approaches based on \emph{state elimination} have been
proposed~\cite{Daws04,param_sttt}. The idea is to replace states
and their incident transitions by direct transitions
from each predecessor to each successor state. Eliminating states this
way iteratively leads to a model having only initial and absorbing
states, where transitions from the initial states to the absorbing states
carry---as rational functions over the model parameters---the
probability of reaching the absorbing states from the initial states.
The efficiency of such elimination methods strongly depends on the
order in which states are eliminated and on the representation of the
rational functions.

\paragraph{Related work}  The idea of constructing a regular expression
representing a DTMC's behavior originates from Daws~\cite{Daws04}. He
uses state elimination to generate regular expressions describing the
paths from the initial states to the absorbing states of a DTMC.  Hahn
\textit{et al.}~\cite{param_sttt} apply this idea to PDTMCs to obtain
rational functions for reachability and expected reward properties.
They improve the efficiency of the construction by heuristics for the
transformation of finite automata to regular
expressions~\cite{Gruber08} to guide the elimination of states.
Additionally, they reduce the polynomials to simplify the rational
functions. These ideas have been extended to Markov decision
processes~\cite{hahn_param_nfm}. The main problem there is that the
reachability probabilities depend on the chosen scheduler to resolve
the nondeterminism. When maximizing or minimizing these probabilities,
the optimal scheduler generally depends on the values of the
parameters. Their algorithms are implemented in PARAM~\cite{PARAM10},
the---to the best of our knowledge---only available tool for computing
reachability probabilities of PDTMCs.  

Several authors have
considered the related problem of parameter synthesis: for which
parameter instances does a given (LTL or PCTL) formula hold? To
mention a few, Han \emph{et al.}~\cite{han-et-al-rtss-08} considered
this problem for timed reachability in continuous-time Markov chains,
Pugelli \emph{et al.}~\cite{seshia_et_al_cav_13} for Markov decision
processes, and Benedikt \emph{et al.}~\cite{benedikt-et-al-tacas-2013}
for $\omega$-regular properties of interval Markov chains.

\paragraph{Contributions of this paper}
In this paper we improve the computation of reachability probabilities for 
PDTMCs~\cite{Daws04,param_sttt} in two important ways. We introduce a state 
elimination strategy based on a \emph{recursive graph decomposition} of the PDTMC into 
strongly connected subgraphs and give a novel method to \emph{efficiently factorize 
polynomials}. Although presented in the context of parametric Markov chains, this 
constitutes a generic method for representing and manipulating rational 
functions and is well-suited for other applications as well. The experiments 
show that using our techniques yield a speed-up of up to
three orders of magnitude compared to~\cite{param_sttt} on many benchmarks. 

  \section{Preliminaries}
\label{sec:preliminaries}

%In this section we brief{}ly introduce the basic concepts of
%discrete-time Markov chains and formalize the extension to
%parametric Markov chains.

\begin{definition}[Discrete-time Markov chain]
\label{def:dtmc}
  A \emph{discrete-time Markov chain (DTMC)} is a tuple $\DTMCdef$ with 
  a non-empty finite set $S$ of states, an initial distribution $I:S\to[0,1]\subseteq\R$ 
  with $\sum_{s\in S}I(s)=1$, and a transition probability matrix 
  $P:S\times S\to[0,1]\subseteq\R$ with $\sum_{s'\in S}P(s,s')=1$ for all $s\in S$.
\end{definition}

The states $\Sinit = \{\sinit\in S\,|\,I(\sinit)>0\}$ are called \emph{initial states}.
A \emph{transition} leads from a state $s\in S$ to a state $s'\in S$ iff $P(s,s')>0$.
The set of \emph{successor states of $s\in S$} is $\mathrm{succ}(s)=\{s'\in S\,|\,P(s,s')>0\}$.
A \emph{path} of $\DTMCD$ is a finite sequence $\pi = s_0 s_1 \ldots s_n$ of states 
$s_i\in S$ such that $P(s_i,s_{i+1})>0$ for all $0\leq i < n$. 
The set $\pathsfin^{\DTMCD}$ contains all paths of ${\DTMCD}$, $\pathsfin^{\DTMCD}(s)$ those
starting in $s\in S$, and $\pathsfin^{\DTMCD}(s,t)$ 
those starting in $s$ and ending in $t$. We generalize this to 
sets $S',S''\subseteq S$ of states by $\pathsfin^{\DTMCD}(S',S'') = \bigcup_{s'\in 
S'}\bigcup_{s''\in S''}\pathsfin^{\DTMCD}(s',s'')$. A state $t$ is 
\emph{reachable} from $s$ iff $\pathsfin^{\DTMCD}(s,t)\neq\emptyset$.

The \emph{probability measure} $\pfin^{\DTMCD}$ for paths 
satisfies%\footnote{To determine reachability probabilities in a DTMC, for the initial states this product is weighted by the initial distribution values.} 
\[
  \pfin^{\DTMCD}(s_0 {\ldots} s_n) =
  \prod_{i=0}^{n{-}1}P(s_i,s_{i+1})
\]
%\rw{You have to take the initial distribution into account for the
%  probability of a path!}
%\eab{No, only at the very end; the above measure should fit for all states, not olny for the initial ones.}
and $\pfin^{\DTMCD}\bigl(\{\pi_1,\pi_2\}\bigr)=\pfin^{\DTMCD}(\pi_1)+\pfin^{\DTMCD}(\pi_2)$
for all $\pi_1,\pi_2\in\pathsfin^{\DTMCD}$ not being the prefix
of each other.  In general, for $R\subseteq\pathsfin^{\DTMCD}$ we
have $\pfin^{\DTMCD}(R) = \sum_{\pi\in R'} \pfin^{\DTMCD}(\pi)$ with
$R' = \{\pi\in R \mid \forall \pi'\in R.\ \pi' \text{ is not a proper
  prefix}$ $\text{of } \pi \}$. We often omit the superscript $\DTMCD$
if it is clear from the context. For more details see, \eg
\cite{BK08}.

For a DTMC ${\DTMCdef}$ and some $K\subseteq S$
we define the set of \emph{input states} of $K$ by $\Inp(K)=\{s\in K\mid
I(s)>0 \vee \exists s'\in S\setminus K.\ P(s',s)>0\}$, \ie the states inside $K$ that
have an incoming transition from outside $K$. Analogously, we define the
set of \emph{output states} of $K$ by $\Out(K)=\{s\in S\setminus K\mid\exists
s'\in K.\ P(s',s)>0\}$, \ie the states outside $K$ that
have an incoming transition from a state inside $K$. The set of \emph{inner
states} of $K$ is given by $K\setminus\Inp(K)$.

We call a state set $S'\subseteq S$ \emph{absorbing} iff there is a
state $s'\in S'$ from which no state outside $S'$ is reachable in ${\DTMCD}$, \ie iff
$\pathsfin^{\DTMCD}(\{s'\},S\setminus S') = \emptyset$. A state $s\in S$ is
absorbing if $\{s\}$ is absorbing.

A set $S'\subseteq S$ induces a \emph{strongly connected subgraph
  (SCS) of ${\DTMCD}$} iff for all $s,t\in S'$ there is a path from
$s$ to $t$ visiting only states from $S'$.  A \emph{strongly connected
  component (SCC)} of ${\DTMCD}$ is a maximal (w.\,r.\,t.
$\subseteq$) SCS of $S$. An SCC $S'$ is called \emph{bottom} if
$\Out(S')=\emptyset$ holds. The probability of eventually reaching a
bottom SCC in a finite DTMC is always $1$~\cite[Chap.~10.1]{BK08}.

We consider \emph{probabilistic reachability properties}, putting
bounds on the probability
$\sum_{\sinit\in \Sinit}I(\sinit)\cdot\pfin^{\DTMCD}\bigl(\pathsfin^{\DTMCD}(\sinit,T)\bigr)$ to eventually
reach a set $T\subseteq S$ of states from the initial
states.  It is well-known that this suffices for checking arbitrary
$\omega$-regular properties, see \cite[Chap.~10.3]{BK08} for the
details.

Note that the probability of reaching a state in a bottom SCC equals
the probability of reaching one of the input states of the bottom SCC.
Therefore, we can make all input states of bottom SCCs absorbing,
without loss of information. Furthermore, if we are interested in the
probability to reach a given state, also this state can be made
absorbing without modifying the reachability probability of interest.
Therefore, in the following we consider only models whose bottom SCCs
are single absorbing states forming the set $T$ of \emph{target}
states, whose reachability probabilities are of interest.

\subsection{Parametric Markov Chains}
\label{sec:pdtmcs}

To add parameters to DTMCs, we follow~\cite{PARAM10} by  
allowing arbitrary rational functions in the definition of probability 
distributions. 
%Let in the following  $\dom(f)=A$ denote the
%\emph{domain} of a function $f:A\rightarrow B$.

\begin{definition}[Polynomial and rational function]
  \label{def:polynomial}
  Let $V=\{\series{x}{n}\}$ be a finite set of \emph{variables} with domain
  $\R$. A \emph{polynomial} $\pol$ over $V$ is a sum of \emph{monomials}, which are
  products of variables in $V$ and a coefficient in $\Z$:
  \[
  \pol=a_1\cdot x_{1}^{e_{1,1}}\cdot\ldots\cdot x_{n}^{e_{1,n}}\ +\ \cdots \ + 
       \ a_m\cdot x_{1}^{e_{m,1}}\cdot\ldots\cdot x_{n}^{e_{m,n}},
  \]
  where $e_{i,j}\in\N_0=\N\cup\{0\}$ and $a_i\in\Z$ for all $1\leq i\leq m$ 
  and $1\leq j\leq n$. $\Z[\series{x}{n}]$ denotes the set of polynomials over
  $V=\{\series{x}{n}\}$.
  A \emph{rational function} over $V$ 
  is a quotient $f=\frac{\pol_1}{\pol_2}$ of two polynomials $\pol_1, \pol_2$ over $V$ with $\pol_2 \neq\nullpoly$\footnote{$\pol_2 \neq \nullpoly$ means that $\pol_2$ cannot be simplified to $\nullpoly$.}. We use
  $\ratfunc = \bigl\{ \frac{\pol_1}{\pol_2} \,|\, \pol_1,\pol_2\in \Z[\series{x}{n}]\land \pol_2\neq\nullpoly\bigr\}$ 
  to denote the set of rational functions over $V$.
\end{definition}
% Using rational functions we are now able to extend the notion of DTMCs
% to their parametric counterpart.
\begin{definition}[PDTMC]
  \label{def:dtpmc}
  A \emph{parametric discrete-time Markov chain (PDTMC)} is a tuple $\PDTMCdef$ 
  with a finite set of states $S$, a finite set of 
  parameters $V=\{\series{x}{n}\}$ with domain $\R$, an initial distribution $I:S \rightarrow \ratfunc$, and a parametric transition 
  probability matrix $P: S \times S \rightarrow \ratfunc$.
\end{definition}

The \emph{underlying graph} $\graph_{\PDTMCM}=(S,\mathcal{D}_P)$ of a
(P)DTMC $\PDTMCdef$ is given by $\mathcal{D}_P=\bigl\{(s,s')\in
S\times S\,\big|\,P(s,s')\neq\nullpoly\bigr\}$. As for DTMCs, we assume
that all bottom SCCs of considered PDTMCs are single absorbing states.

%Using an
%\emph{evaluation}, all or some of the parameters occurring in the
%rational functions of a PDTMC can be instantiated.

\begin{definition}[Evaluated PDTMC]
\label{def:evPDTMC}
  An \emph{evaluation $u$ of $V$} is a function $u\colon V \rightarrow \R$. 
  The evaluation $\evalratfunc{\pol}{V}$ of a polynomial 
  $\pol\in \Z[\series{x}{n}]$ under $u\colon V\rightarrow\R$ substitutes 
  each $x \in V$ by $u(x)$, using the standard semantics for $+$ and $\cdot$. 
  For $f=\frac{\pol_1}{\pol_2} \in \ratfunc$ we define
  $\evalratfunc{f}{V}=\frac{\evalratfunc{\pol_1}{V}}{\evalratfunc{\pol_2}{V}}\in\R$ if $\evalratfunc{\pol_2}{V}\neq 0$.

  For a PDTMC $\PDTMCdef$ and an evaluation $u$, the \emph{evaluated
    PDTMC} is the DTMC $\DTMCudef$ given by $S_u=S$ and for all $s, s'
  \in S_u$, $I_u(s)= \evalratfunc{I(s)}{V}$ and $P_u(s,
  s')=\evalratfunc{P(s, s')}{V}$ if the evaluations are defined and
  $0$ otherwise.
\end{definition}
An evaluation $u$ substitutes each parameter 
by a real number. This induces a well-defined probability measure on the evaluated PDTMC under the following conditions.

\begin{definition}[Well-defined evaluation]
  \label{def:well_evaluation}
  An evaluation $u$ is \emph{well-defined} for a PDTMC $\PDTMCdef$ if for the evaluated PDTMC $\DTMCudef$ it holds that
  \begin{itemize}
  \item $I_u:S_u\to [0,1]$ with $\sum_{s\in S_u}I_u(s) = 1$, and
  \item $P_u:S_u\times S_u\rightarrow[0,1]$ with $\sum_{s' \in S_u}
    P_u(s, s') = 1$ for all $ s \in S_u$.
  \end{itemize}
  An evaluation $u$ is called \emph{graph preserving} if is well-defined and it holds that
  \[
    \forall s, s' \in S: P(s, s') \neq\nullpoly \implies \evalratfunc{P(s,s')}{V}>\nullnum.
  \]
\end{definition}
Note that $\evalratfunc{P(s,s')}{V}>0$ implies that no division by $0$ will occur. This 
will be ensured during the model checking algorithm, requiring the
evaluation $u$ to be graph preserving, \ie $\graph_{\PDTMCM}=\graph_{\DTMCMu}$.
This is necessary, otherwise altering the graph could make reachable states unreachable, thereby changing
reachability probabilities. 
%Taking this into account,
%the probability of reaching target states is again a rational function.

\begin{definition}\label{def:param_model_check}
  Given a PDTMC $\PDTMCdef$ with absorbing states $T\subseteq S$, the
  \emph{parametric probabilistic model checking problem} is to find for
  each initial state $\sinit\in\Sinit$ and each $\starget\in T$ a rational function
  $f_{\sinit,\starget}\in\ratfunc$ such that for all graph-preserving
  evaluations $u:V\rightarrow\R$ and the evaluated PDTMC
  $\DTMCudef$ it holds that 
  $\evalratfunc{f_{\sinit,\starget}}{V}\ = \ \pr^{\PDTMCM_u}\bigl(\paths^{\PDTMCM_u}(\sinit,\starget)\bigr)$. 
\end{definition}

Given the functions $f_{\sinit,\starget}$ for $\sinit\in\Sinit$ and $\starget\in T$, 
the probability of reaching a state in $T$ from an initial state is
$\sum_{\sinit\in\Sinit}I(\sinit)\cdot\Bigl(\sum_{\starget\in T} f_{\sinit,\starget}\Bigr)$.

  \section{Parametric Model Checking by SCC Decomposition}
\label{sec:scc_reduction}
In this section we present our algorithmic approach to apply model
checking to PDTMCs. In the following let $\PDTMCdef$ be a PDTMC
with absorbing state set
$T\subseteq S$.
 For each initial state $\sinit\in\Sinit$ and each target state
$\starget\in T$ we compute a rational function $f_{\sinit,\starget}$ over the
set of parameters $V$ which describes the probability of reaching $\starget$
from $\sinit$ as in~\cite{param_sttt}. 
We do this using \emph{hierarchical graph decomposition}, inspired by
a former method for computing reachability probabilities in the
non-parametric case~\cite{AJWKB10}.

\subsection{PDTMC Abstraction}

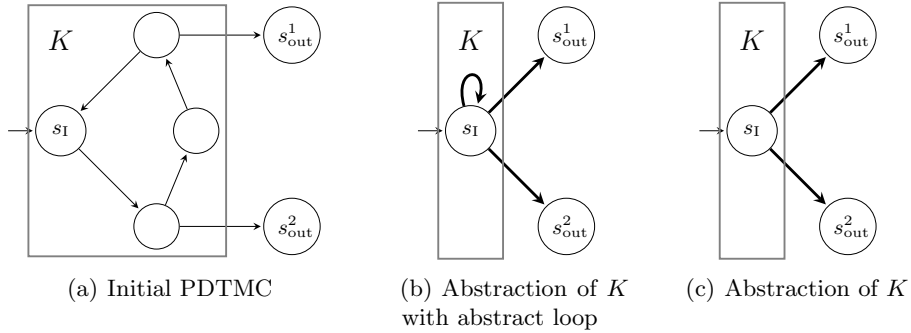
\begin{figure}[t]
  \begin{center}    
    \subfigure[Initial PDTMC]
	{%
	  \scalebox{0.9}{\begin{tikzpicture}[>=stealth,shorten >= 1pt,text centered,text width=12pt,node distance=2cm]
  \draw[white, use as bounding box] (-1,-2) rectangle (4,2.1);

  \node[draw, circle, text centered] at (0,0) (sinit) {$\sinit$};  
  \node[draw, circle, text centered] [on grid, below right of=sinit] (1) {};
  \node[draw, circle, text centered] [on grid, above right of=sinit] (2) {};
  \node[draw, circle, text centered] [on grid, right of=sinit] (3) {};
  \node[draw, circle, text centered] [on grid, right of=1] (sout1) {$s_{\mathrm{out}}^2$};
  \node[draw, circle, text centered] [on grid, right of=2] (sout2) {$s_{\mathrm{out}}^1$};

  \node [left=.4cm of sinit] (dummy) {};  
  \node [above=0.7cm of sinit] (k) {\large$K$};

\path[->]
  (dummy) edge [] (sinit)
;
\path[->]
  (sinit) edge[] (1)
  (1) edge[] (3)
  (3) edge[] (2)
  (2) edge[] (sinit)
  (1) edge[] (sout1)
  (2) edge[] (sout2)
;
\node [rectangle] (box) [draw=grey, thick, fit = (sinit) (1) (2) (3),  inner sep = 0.1cm] {};

\end{tikzpicture}}%
	  \label{fig:pdtmc_abstraction_K}
	}\hfill%    
     \subfigure[Abstraction of $K$ with abstract loop]
	{%
	  \scalebox{0.9}{\begin{tikzpicture}[>=stealth,shorten >= 1pt,text centered,text width=12pt,node distance=2cm]
  \draw[white, use as bounding box] (-1,-2) rectangle (2.1,2.1);

  \node[draw, circle, text centered] at (0,0) (sinit) {$\sinit$};  
  \phantom{%
  \node[draw, circle, text centered] [on grid, below=1.45cm of sinit] (1) {};
  \node[draw, circle, text centered] [on grid, above=1.45cm of sinit] (2) {};
  }%
  \node[draw, circle, text centered] [on grid, above right of=sinit] (sout1) {$s_{\mathrm{out}}^1$};
  \node[draw, circle, text centered] [on grid, below right of=sinit] (sout2) {$s_{\mathrm{out}}^2$};
  
  \node [left=.4cm of sinit] (dummy) {};  
  \node [above=0.7cm of sinit] (k) {\large$K$};

\path[->]
  (dummy) edge [] (sinit)
;
\path[->]
  (sinit) edge[very thick] (sout1)
  (sinit) edge[very thick] (sout2)  
;
\path[->]
  (sinit) edge [loop above, very thick] (sinit);

\node [rectangle] (box) [draw=grey, thick, fit = (sinit) (1) (2),  inner sep = 0.1cm] {};

\end{tikzpicture}}%
	  \label{fig:pdtmc_abstraction_abstract_loop}
	}\hfill%     
     \subfigure[Abstraction of $K$]
	{%
	  \scalebox{0.9}{\begin{tikzpicture}[>=stealth,shorten >= 1pt,text centered,text width=12pt,node distance=2cm]
  \draw[white, use as bounding box] (-1,-2) rectangle (2.1,2.1);

  \node[draw, circle, text centered] at (0,0) (sinit) {$\sinit$};  
  \phantom{%
  \node[draw, circle, text centered] [on grid, below=1.45cm of sinit] (1) {};
  \node[draw, circle, text centered] [on grid, above=1.45cm of sinit] (2) {};
  }%
  \node[draw, circle, text centered] [on grid, above right of=sinit] (sout1) {$s_{\mathrm{out}}^1$};
  \node[draw, circle, text centered] [on grid, below right of=sinit] (sout2) {$s_{\mathrm{out}}^2$};
  
  \node [left=.4cm of sinit] (dummy) {};  
  \node [above=0.7cm of sinit] (k) {\large$K$};

\path[->]
  (dummy) edge [] (sinit)
;
\path[->]
  (sinit) edge[very thick] (sout1)
  (sinit) edge[very thick] (sout2)  
;
\node [rectangle] (box) [draw=grey, thick, fit = (sinit) (1) (2),  inner sep = 0.1cm] {};

\end{tikzpicture}}%
	  \label{fig:pdtmc_abstraction_abstract}
	}%    
  \end{center}%
 \caption{The concept of PDTMC abstraction}
 \label{fig:pdtmc_abstraction}
\end{figure}

The basic concept of our model checking approach is to replace a
non-absorbing subset $K\subseteq S$ of states and all transitions
between them by transitions directly leading from the input states
$\Inp(K)$ of $K$ to the output states $\Out(K$) of $K$, carrying
the accumulated probabilities of all paths between the given input and
output states inside $K$. This concept is illustrated in
Figure~\ref{fig:pdtmc_abstraction}: In
Figure~\ref{fig:pdtmc_abstraction_K}, 
$K$ has one input state $\sinit$ and two output states
$s_{\mathrm{out}}^1$, $s_{\mathrm{out}}^2$. The abstraction in
Figure~\ref{fig:pdtmc_abstraction_abstract} hides every state of $K$
except for $\sinit$; all transitions are directly leading to the
output states.

As we need a probability measure for arbitrary subsets of states, we first 
define sub-PDTMCs induced by such subsets.
\begin{definition}[Induced PDTMC]
\label{def:inducedPDTMC}
Given a PDTMC \PDTMCdef and a non-absorbing subset $K \subseteq S$ of states,
the \emph{PDTMC induced by $\PDTMCM$ and $K$} is given by
$\PDTMCM^K=(S^K,V^K,I^K,P^K)$ with $S^K = K \cup \Out(K)$, $V^K = V$, 
and for all $s,s' \in S^K$,  $I^K(s) \neq 0 \iff s \in \Inp(K)$ and
  \[
  P^K(s,s') = \begin{cases}
                          P(s,s'), & \text{if $s \in K, s' \in S^K$,} \\
                          1, & \text{if $s=s' \in \Out(K)$,} \\
                          0, & \text{otherwise.}
               \end{cases}
 \]
\end{definition}

Intuitively, all incoming and outgoing transitions are preserved for inner
states of $K$ while the output states are made absorbing. We allow an arbitrary
input distribution $I^K$ with the only constraint that $I^K(s)\neq 0$ iff $s$
is an input state of $K$.

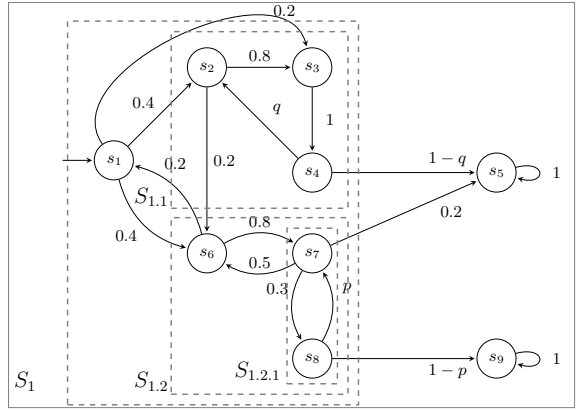
\begin{figure}[tb]
  \centering
  \scalebox{0.7}{\begin{tikzpicture}[>=stealth,shorten >= 1pt,text centered,text width=12pt,node distance=2cm]
  \draw[grey, use as bounding box] (-2,-8.7) rectangle (8.75,-1);

  \node[draw, circle, text centered] at (0, -4) (1) {$s_1$};
  \node[draw, circle, text centered] [on grid, above right=2.5cm of 1] (2) {$s_2$};
  \node[draw, circle, text centered] [on grid, right of=2] (3) {$s_3$};
  \node[draw, circle, text centered] [on grid, below of=3] (4) {$s_4$};
  \node[draw, circle, text centered] [on grid, right=3.5cm of 4] (5) {$s_5$};
  \node[draw, circle, text centered] [on grid, below right=2.5cm of 1] (6) {$s_6$};
  \node[draw, circle, text centered] [on grid, right of=6] (7) {$s_7$};
  \node[draw, circle, text centered] [on grid, below of=7] (8) {$s_8$};
  \node[draw, circle, text centered] [on grid, right=3.5cm of 8] (9) {$s_9$};
  \node [on grid, left=1.3cm of 1] (initialphantom) {};

\path[->]
  (initialphantom) edge (1)
  (1) edge[above left] node {$0.4$} (2)
  (1) edge[bend left=100, above] node[pos=0.85] {$0.2$} (3)
  (1) edge[below left, bend right] node {$0.4$} (6)
  (2) edge[above] node {$0.8$} (3)
  (2) edge[right] node {$0.2$} (6)
  (3) edge[right] node {$1$} (4)
  (4) edge[above right] node {$q$} (2)
  (4) edge[above] node[near end] {${1-q}$} (5)
  (5) edge[above, loop right] node {$1$} (5)
  (6) edge[above right, bend right] node[near end] {$0.2$} (1)
  (6) edge[bend left,above] node {$0.8$} (7)
  (7) edge[above left] node[below right,pos=0.7] {$0.2$} (5)
  (7) edge[bend left,above] node {$0.5$} (6)
  (7) edge[left, near start, bend right] node {$0.3$} (8)
  (8) edge[right, near end, bend right] node {$p$} (7)
  (8) edge[below] node[near end] {${1-p}$} (9)
  (9) edge[above, loop right] node {$1$} (9)
;

\node [rectangle] (scc1) [draw=grey, thick, dashed, fit = (1) (3) (8), inner sep =0.5cm] {};
\node [rectangle] (scc1.1) [draw=grey, thick, dashed, fit = (2) (4), inner sep = 0.3cm] {};
\node [rectangle] (scc1.2) [draw=grey, thick, dashed, fit = (6) (8), inner sep = 0.3cm] {};
\node [rectangle] (scc1.2.1) [draw=grey, thick, dashed, fit = (7) (8), inner sep = 0.1cm] {};

\node at (-1.7,-8.2) (s1) {\large{$S_1$}};
\node at (0.6,-4.7) (s1.1) {\large{$S_{1.1}$}};
\node at (0.6,-8.2) (s1.2) {\large{$S_{1.2}$}};
\node at (2.5,-8) (s1.2.1) {\large{$S_{1.2.1}$}};

\end{tikzpicture}}
  \caption{Example PDTMC and its SCC decomposition}
  \label{fig:example_pdtmc}
\end{figure}

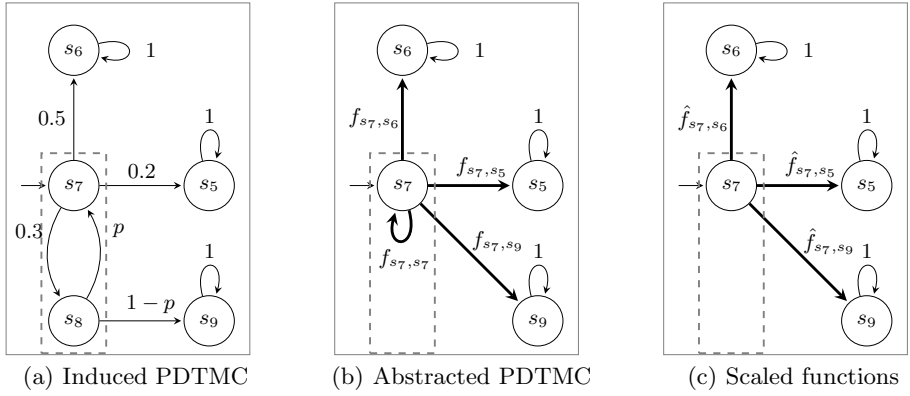
\begin{figure}[t]
  \begin{center}    
    \subfigure[Induced PDTMC]
	{%
	  \scalebox{0.9}{\begin{tikzpicture}[>=stealth,shorten >= 1pt,text centered,text width=12pt,node distance=2cm]
  \draw[grey, use as bounding box] (-1,-4.5) rectangle (2.6,.7);

  \node[draw, circle, text centered] at (0,0) (6) {$s_6$};  
  \node[draw, circle, text centered] [on grid, below of=6] (7) {$s_7$};
  \node[draw, circle, text centered] [on grid, right of=7] (5) {$s_5$};
  \node[draw, circle, text centered] [on grid, below of=7] (8) {$s_8$};
  \node[draw, circle, text centered] [on grid, below of=5] (9) {$s_9$};
  
  \node [left=.4cm of 7] (dummy) {};  

\path[->]
  (dummy) edge [] (7)
  (5) edge[above, loop above] node {$1$} (5)
  (7) edge[above left] node[above] {$0.2$} (5)
  (7) edge[above] node [left] {$0.5$} (6)
  (7) edge[left, near start, bend right] node {$0.3$} (8)
  (8) edge[right, near end, bend right] node {$p$} (7)
  (8) edge[below] node[above] {${1-p}$} (9)
  (9) edge[above, loop above] node {$1$} (9)
  (6) edge[above, loop right] node {$1$} (6)
;

\node [rectangle] (scc1.2.1) [draw=grey, thick, dashed, fit = (7) (8), inner sep = 0.1cm] {};

\end{tikzpicture}}%
	  \label{fig:induced_pdtmc_ex}
	}\hfill%    
	\subfigure[Abstracted PDTMC]
	{%
	  \scalebox{0.9}{\begin{tikzpicture}[>=stealth,shorten >= 1pt,text centered,text width=12pt,node distance=2cm]
  \draw[grey, use as bounding box] (-1,-4.5) rectangle (2.6,.7);

  \node[draw, circle, text centered] at (0,0) (6) {$s_6$};  
  \node[draw, circle, text centered] [on grid, below of=6] (7) {$s_7$};
  \node[draw, circle, text centered] [on grid, right of=7] (5) {$s_5$};
  \phantom{\node[draw, circle, text centered] [on grid, below of=7] (8) {$s_8$};}
  \node[draw, circle, text centered] [on grid, below of=5] (9) {$s_9$};
  
  \node [left=.4cm of 7] (dummy) {};  

\path[->]
  (dummy) edge [] (7)
  (5) edge[above, loop above] node {$1$} (5)
  (9) edge[above, loop above] node {$1$} (9)
;
\path[->]
  (7) edge[above left, very thick] node[above] {$f_{s_7,s_5}$} (5)
  (7) edge[above, very thick] node [left=7pt] {$f_{s_7,s_6}$} (6)
  (7) edge[below, very thick] node[right=-3pt,yshift=5pt] {$f_{s_7,s_9}$} (9)
  (7) edge[very thick,loop below] node[below,xshift=-4pt] {$f_{s_7,s_7}$} (7)  
  (6) edge[above, loop right] node {$1$} (6)
;

\node [rectangle] (scc1.2.1) [draw=grey, thick, dashed, fit = (7) (8), inner sep = 0.1cm] {};

\end{tikzpicture}}%
	  \label{fig:pdtmc_abstraction_ex}
	}\hfill%
	\subfigure[Scaled functions]
	{%
	  \scalebox{0.9}{\begin{tikzpicture}[>=stealth,shorten >= 1pt,text centered,text width=12pt,node distance=2cm]
  \draw[grey, use as bounding box] (-1,-4.5) rectangle (2.6,.7);

  \node[draw, circle, text centered] at (0,0) (6) {$s_6$};  
  \node[draw, circle, text centered] [on grid, below of=6] (7) {$s_7$};
  \node[draw, circle, text centered] [on grid, right of=7] (5) {$s_5$};
  \phantom{\node[draw, circle, text centered] [on grid, below of=7] (8) {$s_8$};}
  \node[draw, circle, text centered] [on grid, below of=5] (9) {$s_9$};
  
  \node [left=.4cm of 7] (dummy) {};  

\path[->]
  (dummy) edge [] (7)
  (5) edge[above, loop above] node {$1$} (5)
  (9) edge[above, loop above] node {$1$} (9)
;
\path[->]
  (7) edge[above left, very thick] node[above] {$\hat{f}_{s_7,s_5}$} (5)
  (7) edge[above, very thick] node [left=7pt] {$\hat{f}_{s_7,s_6}$} (6)
  (7) edge[below, very thick] node[right=-2pt,yshift=5pt] {$\hat{f}_{s_7,s_9}$} (9)
  (6) edge[above, loop right] node {$1$} (6)

;

\node [rectangle] (scc1.2.1) [draw=grey, thick, dashed, fit = (7) (8), inner sep = 0.1cm] {};

\end{tikzpicture}}%
	  \label{fig:pdtmc_abstraction_scaled_ex}
	}%    
  \end{center}%
 \caption{PDTMC Abstraction}
 \label{fig:pdtmc_abstraction_example}
\end{figure}

\begin{example}
  Consider the PDTMC $\PDTMCM$ in Figure~\ref{fig:example_pdtmc}
and the state set $K=\{s_7,s_8\}$ with input states
$\Inp(K)=\{s_7\}$ and output states
$\Out(K)=\{s_5,s_6,s_9\}$. The PDTMC
$\PDTMCM^K=(S^K,V^K,I^K,P^K)$ induced by $\PDTMCM$ and $K$ is shown in
Figure~\ref{fig:induced_pdtmc_ex}.
\end{example} 
Note that, since $K$ is non-absorbing, the probability of eventually
reaching one of the output states is $1$.  The probability of reaching
an output state $t$ from an input state $s$ is determined by the
accumulated probability of all paths $\paths(s,t)$ from $s$ to $t$.
Those paths are composed by a (possibly empty) prefix looping on $s$
and a postfix leading from $s$ to $t$ without returning back to $s$.
In our abstraction this is reflected by abstracting the prefixes by an
abstract self-loop on $s$ with probability $f_{s,s}$ and the postfixes by abstract transitions
from the input states $s$ to the output states $t$ with probability $f_{s,t}$ (see
Figure~\ref{fig:pdtmc_abstraction_abstract_loop}).  
If all loops in $K$ are loops on $s$ then $f_{s,t}$ can be easily computed 
as the sum of the probabilities of all loop-free paths from $s$ to $t$.
In the final
abstraction shown in Figure~\ref{fig:pdtmc_abstraction_abstract}, we
make use of the fact that all paths from $s$ to $t$ can be extended
with the same loops on $s$ as a prefix. Therefore we do not need to
compute the probability of looping on $s$, but can scale the probabilities
$f_{s,t}$ such that they sum up to $1$.

\begin{definition}[Abstract PDTMC]
\label{def:abstractPMC}
Let $\PDTMCdef$ be a PDTMC with absorbing states $T\subseteq S$.
The \emph{abstract PDTMC $\PDTMCM\abs=(S\abs,V\abs,I\abs,P\abs)$} is given by
$S\abs = \{s\in S\mid I(s)\neq 0 \lor s\in T\}$, $V\abs = V$, and for all  $s,s' \in S\abs$ we define $I\abs(s) = I(s)$ and 
\[
 P\abs(s, s') = 
    \begin{cases}
      \dfrac{p\abs^{\PDTMCM}(s, s')}{\sum_{s''\in T} p\abs^{\PDTMCM}(s,s'')}, &\text{if $I(s)>0\land s'\in T$,} \\
      1, &\text{if $s=s'\in T$,} \\
      0, &\text{otherwise.}
    \end{cases}
    \]
with
 \[
    p\abs^{\PDTMCM}(s,s')=\pr^{\PDTMCM}\bigl(\{\pi=s_0\ldots s_n \in
    \pathsfin^{\PDTMCM}(s,s')\,|\,s_i\neq s\land s_i\neq s',0<i<n\}\bigr).
 \]
\end{definition}

\begin{example}
  Consider the PDTMC $\PDTMCM'=(S',V',I',P')$ of
  Figure~\ref{fig:induced_pdtmc_ex} with initial state $s_7$ and
  target states $T'=\{s_5,s_6,s_9\}$. The first abstraction step regarding the probabilities
  $p\abs^{\PDTMCM}(s,s')$ is depicted in
  Figure~\ref{fig:pdtmc_abstraction_ex} and refers to the following
  probabilities:
\begin{align*}
  f_{s_7,s_5}\ =\ p\abs^{\PDTMCM'}(s_7,s_5) &=\ 0.2 &  f_{s_7,s_6}\ =\ p\abs^{\PDTMCM'}(s_7,s_6) &=\ 0.5\\
  f_{s_7,s_7}\ =\ p\abs^{\PDTMCM'}(s_7,s_7) &=\ 0.3\cdot p & f_{s_7,s_9}\ =\ p\abs^{\PDTMCM'}(s_7,s_9) &=\ 0.3\cdot (1-p)
\end{align*}
The total probabilities of reaching the output states in $\PDTMCM'\abs$ are given by paths which
first use the loop on $s_7$ arbitrarily many times (including zero times) and
then take a transition to an output state. For example, using the geometric
series, the probability of the set of paths leading from $s_7$ to $s_5$ is given
by
\[
\sum_{i=0}^\infty (f_{s_7,s_7})^i\cdot f_{s_7,s_5}\ =\ \dfrac{1}{1-f_{s_7,s_7}}\cdot f_{s_7,s_5}\, .
\]
As the probability of finally reaching the set of absorbing states in $\PDTMCM'$
is $1$, we can directly scale the probabilities of the outgoing edges such that
their sum is equal to $1$. This is achieved by dividing each of these probabilities by the sum of all
probabilities of outgoing edges, $f_{\mathrm{out}}=0.2+0.5+0.3\cdot (1-p) = 1 - 0.3p$. 

Thus the abstract PDTMC
  $\PDTMCM'\abs=(S'\abs,V'\abs,I'\abs,P'\abs)$ depicted in  Figure~\ref{fig:pdtmc_abstraction_scaled_ex} has states
  $S'\abs=\{s_5,s_6,s_7,s_9\}$ and edges from $s_7$ to all other
  states with the following probabilities:
\begin{align*}
  \hat{f}_{s_7,s_5}\ &=\ 0.2\ / f_{\mathrm{out}} & \hat{f}_{s_7,s_6}\ &=\ 0.5\ / f_{\mathrm{out}}\\
  \hat{f}_{s_7,s_9}\ &=\ \bigl(0.3\cdot (1-p)\bigr)\ / f_{\mathrm{out}}
\end{align*}
\end{example}

\newcommand{\pdtmcabstractiontheorem}{%
  Assume a PDTMC $\PDTMCdef$ with absorbing states $T\subseteq S$, and
  let $\PDTMCM\abs$ be the abstraction of $\PDTMCM$. Then for
  all $\sinit\in\Sinit$ and $\starget\in T$ it holds that
      \[  
    \pfin^{\PDTMCM}\bigl(\pathsfin^{\PDTMCM}(\sinit,\starget)\bigr)=\pfin^{\PDTMCM\abs}\bigl(\pathsfin^{\PDTMCM\abs}(\sinit,\starget)\bigr)\ .
      \]
}
\begin{theorem}\label{theo:abstraction}
  \pdtmcabstractiontheorem
\end{theorem}
The proof of this theorem can be found in the appendix. 
It remains to define the substitution of subsets of states by their
abstractions. Intuitively, a subset of states is replaced by the abstraction as
in Definition~\ref{def:abstractPMC}, while the incoming transitions of the
initial states of the abstraction as well as the 
outgoing transitions of the
absorbing states of the abstraction remain unmodified.
\begin{definition}[Substitution]
\label{def:substitution}
 Assume a PDTMC $\PDTMCdef$, a non-absorbing set $K\subseteq S$ of states, the induced PDTMC 
$\PDTMCM^K=(S^K,V^K,I^K,P^K)$ and the abstraction $\PDTMCM^K\abs=(S^K\abs,V^K\abs,I^K\abs,P^K\abs)$. The 
\emph{substitution of $\PDTMCM^K$ by its abstraction $\PDTMCM^K\abs$ in $\PDTMCM$} is given by
$\PDTMCM\kabs=(S\kabs,V\kabs,I\kabs,P\kabs)$ with $S\kabs= (S\setminus K)\cup S\abs^K$, $V\kabs= V$ and for all $s,s'\in S\kabs$, $I\kabs(s)=I(s)$ and
\[
P\kabs(s,s')=
    \begin{cases}
      P(s,s'),       &\mbox{if } s\notin K,\\
      P^K\abs(s,s'), &\mbox{if } s\in K\land s'\in\Out(K),\\
      0,             &\mbox{otherwise}.
    \end{cases}
\]
\end{definition}
Due to Theorem~\ref{theo:abstraction}, it directly follows that this
substitution does not change reachability properties from the initial states to the
absorbing states of a PDTMC.
\begin{corollary}
  Given a PDTMC $\PDTMCM$ and a non-absorbing subset $K\subseteq S$ of states, it holds for
all initial states $\sinit\in\Sinit$ and absorbing states $\starget\in T$ that
  \[
   \pfin^{\PDTMCM}\bigl(\pathsfin^{\PDTMCM}(\sinit,\starget)\bigr)=\pfin^{\PDTMCM\kabs}\bigl(\pathsfin^{\PDTMCM\kabs}(\sinit,\starget)\bigr)\,.
  \]
\end{corollary}
\subsection{Model Checking Parametric Markov Chains}
In the previous section we gave the theoretical background for our
model checking algorithm. Now we describe how to compute the abstractions efficiently.

\begin{algorithm}[tb]
\caption{Model Checking PDTMCs}
\label{algo:model_check}
  \begin{newalgorithm}{abstract}{PDTMC $\PDTMCM$}
    \ForAll {non-bottom SCCs $K$ in $\PDTMCM^{S\setminus\Inp(\PDTMCM)}$}\label{algo:model_check:scc_search}
    \State $\PDTMCM^K\abs\colonequals$ abstract($\PDTMCM^K$)\label{algo:model_check:call_rek}\\
    \State $\PDTMCM\colonequals\PDTMCM\kabs$\label{algo:model_check:substitute}\\
    \EndFor
    \State $K\colonequals \{\textit{non-absorbing states in } \PDTMCM\}$ \\
    \State $\PDTMCM\colonequals\PDTMCM_{K\mapsto \mathrm{abs}}$ \\
    \Return{$\PDTMCM$}\label{algo:model_check:return_abs}
    \setcounter{oldlinenumber}{\value{linenno}}
  \end{newalgorithm}
  \begin{newalgorithm}{model\_check}{PDTMC $\PDTMCdef$, $T\subseteq \{t\in S\,|\, P(t,t)=1\}$}
      \setcounter{linenno}{\value{oldlinenumber}}
      \State $\PDTMCM\abs=(S\abs,V\abs, I\abs,P\abs)\colonequals$ abstract($\PDTMCM$)\label{algo:model_check:compute_abs}\\
      \Return{$\sum\limits_{\sinit\in\Sinit}I(\sinit)\cdot\Bigl(\sum\limits_{t\in T}P\abs(\sinit,t)\Bigr)$}\label{algo:model_check:mc_result}
  \end{newalgorithm}
\end{algorithm}

As a heuristic for forming the sets of states to be abstracted, we
choose an SCC-based decomposition of the graph.  Algorithmically,
Tarjan's algorithm~\cite{Tarjan72} is used to determine the SCC
structure of the graph while we do not consider bottom SCCs. We
hierarchically determine also sub-SCCs inside the SCCs without their input
states, until no non-trivial sub-SCCs exist any more.

\begin{example} In
Figure~\ref{fig:example_pdtmc}, the dashed rectangles indicate the
decomposition into the SCC $S_1=\{1,2,3,4,6,7,8\}$ and the sub-SCSs
$S_{1.1}=\{2,3,4\}$, $S_{1.2}=\{6,7,8\}$, and
$S_{1.2.1}=\{7,8\}$ with
$S_{1.1}\subset S_1$ and $S_{1.2.1}\subset
S_{1.2} \subset S_1$. 
\end{example}
 
The general model checking algorithm is depicted in
Algorithm~\ref{algo:model_check}. The recursive method
\textit{abstract}(PDTMC $\PDTMCM$) computes the abstraction
$\PDTMCM\abs$ by iterating over all SCCs of the graph without
the input states of $\PDTMCM$
(line~\ref{algo:model_check:scc_search}). For each SCC $K$, the
abstraction $\PDTMCM^K\abs$ of the induced PDTMC $\PDTMCM^K$ is
computed by a recursive call of the method
(line~\ref{algo:model_check:call_rek},
Definitions~\ref{def:inducedPDTMC},\ref{def:abstractPMC}). Afterwards,
$\PDTMCM^K$ is substituted by its abstraction inside $\PDTMCM$
(line~\ref{algo:model_check:substitute},
Definition~\ref{def:substitution}). Finally, the abstraction
$\PDTMCM\abs$ is computed and returned
(line~\ref{algo:model_check:return_abs},
Definition~\ref{def:abstractPMC}).  This method is called by the model
checking method (line~\ref{algo:model_check:compute_abs}) which yields
the abstract system $\PDTMCM\abs$, in which transitions lead only from
the initial states to the absorbing states. All transitions are
labeled with a rational function for the reachability probability, as
in Definition~\ref{def:param_model_check}. Then the whole reachability
probability is computed by building the sum of these transitions
(line~\ref{algo:model_check:mc_result}).
%This is compared to the given upper probability bound $\lambda\in \Q$ returning a truth-value. 
%Note that this can be adapted for lower or strict probability bounds as well.
%\rw{Compare the comment below the Algorithm.}

What remains to be explained is the computation of the abstract probabilities
$p\abs^{\PDTMCM}$. We distinguish the cases 
where the set $K$ has one or multiple input states.

\paragraph{One input state}

%We define the set of paths $\Rloop$ going from $\sinit$ to $\sinit$ and 
%the set of paths $\Rout$ going from $\sinit$ to some $\starget\in T$ without revisiting $\sinit$:
%\begin{align}
%  \Rloop= {} &\{\sinit \seriesPath{s}{n} \sinit \in \pathsfin^{\PDTMCM}\ |\
%  \forall 1 \leq i \leq n.\, s_i\notin \{\sinit\}\cup T\}, \label{eq:rloop}\\
%  \Rout\ = {} &\{\sinit \seriesPath{s}{n} \starget \in \pathsfin^{\PDTMCM}\ |\
%  \starget\in T\land \forall 1\leq i\leq n.\,s_i\notin \{\sinit\}\cup T\}.\label{eq:rout}
%\end{align}
Consider a PDTMC $\PDTMCM^K$ induced by $K$ with one initial state
$\sinit$ and the set of absorbing states
$T=\{\starget^1,\ldots,\starget^n\}$, such that
$K\setminus\{\sinit\}$ has no non-trivial SCCs. 
If there is only one absorbing state, \ie $n=1$, we have
$p^{\PDTMCM^K}\abs(\sinit,\starget^1)=1$. This is directly exploited without
further computations.

Otherwise we
determine the probabilities $p\abs^{\PDTMCM^K}(\sinit, \starget^i)$
for all $1 \leq i \leq n$. As $K \setminus \{\sinit\}$ has no non-trivial
SCSs, the set of those paths from $\sinit$ to $\starget^i$ that do not return to $\sinit$ consists of finitely many loop-free
paths. The probability is computed recursively for all
$s\in S^K$ by:
\begin{align}\label{eq:pabs_computation}
  p^{\PDTMCM^K}\abs(s, \starget^i) =
    \begin{cases}
      1, &\text{if $s = \starget^i$,}\\
      \sum\limits_{s'\in (\mathrm{succ}(s)\cap K)\setminus \Inp(K)}P^K(s,s')\cdot p^{\PDTMCM^K}\abs(s',\starget^i), 
         &\text{otherwise.}
    \end{cases}
\end{align}
These probabilities can also be computed by direct or indirect methods for solving 
linear equation systems, see, \eg \cite[Chapters 3,4]{quarteroni_numerical}. 
Note that state elimination as in~\cite{param_sttt} can be applied 
here, too. 
%We tried both while the state elimination approach performed slightly better
%as there are no loops and the number of states is 
%relatively small for these restricted sets.

The probabilities of the abstract PDTMC 
$\PDTMCM^K\abs=(S\abs,V\abs,I\abs,P\abs)$ as in Definition~\ref{def:abstractPMC} 
can now directly be computed, while an additional constraint is added in order to 
avoid divisions by zero:
\begin{align}\label{eq:nozero}
  P^{\PDTMCM^K}\abs(\sinit,\starget^i)=
\begin{cases}
  \frac{p\abs^{\PDTMCM^K}(\sinit,\starget^i)}{\sum_{j=1}^n p\abs^{\PDTMCM^K}(\sinit,\starget^j)}, 
    & \text{if \ $\sum_{j=1}^n p\abs^{\PDTMCM^K}(\sinit,\starget^j)\neq 0$,}\\
  0, & \text{otherwise.}
\end{cases}
\end{align}

\paragraph{Multiple input states}
Given a PDTMC $\PDTMCM^K$ with initial states
$S_I=\{\sinit^1,\ldots,\sinit^m\}$, $m>1$, such that $I^K(\sinit^i)>0$ for all $1\leq i \leq
m$, and absorbing states $T=\{\starget^1,\ldots,\starget^n\}$. The
intuitive idea would be to maintain a copy of $\PDTMCM^K$ for each initial state and
handle the other initial states as inner states in this copy. Then, the method
as described in the previous paragraph can be used. However, this would be expensive in terms of both 
time and memory. Therefore, we first formulate the linear equation
system as in Equation~\eqref{eq:pabs_computation}. All variables
$p^{\PDTMCM^K}\abs(s,t^i)$ with $s\in K\setminus \Inp(K)$ are eliminated from
the equation system. Then for each initial state $\sinit^ i$
the equation system is solved separately by eliminating all 
variables $p^{\PDTMCM^K}\abs(\sinit^j,t^k)$, $j\not= i$. 
\medskip

Algorithm~\ref{algo:model_check} returns the rational functions $P^{\PDTMCM^K}\abs(\sinit,\starget)$ for all $\sinit\in S_I$ and $t\in T$ as in Equation~\eqref{eq:nozero}. To allow only graph-preserving evaluations
of the parameters, we perform preprocessing where conditions are collected
according to Definition~\ref{def:well_evaluation} as well as the ones from
Equation~\eqref{eq:nozero}. These constraints can be evaluated by a \emph{SAT-modulo-
theories} (\emph{SMT}) solver for non-linear real arithmetic~\cite{demoura_nlsat}.  
In case the solver returns an evaluation which satisfies the resulting
constraint set, the reachability property is satisfied. Otherwise, the property is violated.

  \section{Factorization of Polynomials}
\label{sec:factorization}

Both the SCC-based procedure as introduced in the last section as well as mere state-elimination~\cite{param_sttt} build rational functions representing reachability probabilities. These rational functions might grow rapidly in both algorithms and thereby form one of the major bottlenecks of this methodology. As already argued in~\cite{param_sttt}, the best way to stem this blow-up is the cancellation of the rational functions in every computation step, which involves---apart from \emph{addition}, \emph{multiplication}, and \emph{division} of rational functions---the rather expensive calculation of the \emph{greatest common divisor} ($\cgcd$) of two polynomials. 

In this section we present a new way of handling this problem: An additional maintenance and storage of (partial) polynomial factorizations can lead to remarkable speed-ups in the $\cgcd$ computation, especially when dealing with symmetrically structured benchmarks where many similar polynomials occur. We present an optimized algorithm called $\gcd$ which \emph{operates on the (partial) factorizations} of the polynomials to compute their $\cgcd$. During the calculations, the factorizations are also refined. On this account we reformulate the arithmetic operations on rational functions such that they preserve their numerator's and denominator's factorizations, if it is possible with reasonable effort.

\paragraph{Factorizations.} In the following we assume that
polynomials are \emph{normalized},  that is they are of the form
  $\pol=a_1\cdot x_{1}^{e_{1,1}}\cdot\ldots\cdot x_{n}^{e_{1,n}}\ +\ \cdots \ +
  \ a_m\cdot x_{1}^{e_{m,1}}\cdot\ldots\cdot x_{n}^{e_{m,n}}$
with $(e_{j,1}, \ldots, e_{j,n})\neq (e_{k,1}, \ldots, e_{k,n})$ 
for all $j,k\in\{1, \ldots, m\}$ with $j\neq k$
and the monomials are ordered, \eg according to the
reverse lexicographical ordering. %~\cite{Book_Basu_RAAlgorithms}.

\begin{definition}[Factorization]
  \label{def:factorization}
  A \emph{factorization} $\factor{\pol}=\{\pol_1^{e_1},\ldots,\pol_n^{e_n}\}$ of a polynomial
  $\pol\neq 0$ is a non-empty set\footnote{We represent a factorization of a
  polynomial as a set; however, in the implementation we use a more
  efficient binary search tree instead.}  of \emph{factors}
  $\pol_i^{e_i}$, where the bases $\pol_i$ are pairwise
  different polynomials and the exponents are $e_i\in\N$ such that $\pol=\prod_{i=1}^n \pol_i^{e_i}$. 
  We additionally set $\factor{0} = \emptyset$.
\end{definition}

For polynomials $\pol, \altpol$ and a factorization
$\factor{\pol}=\{\pol_1^{e_1},\ldots,\pol_n^{e_n}\}$ of $\pol$ let
$\bases{\factor{\pol}}=\{\pol_1,\ldots,\pol_n\}$ and $\exp(\altpol,
\factor{\pol})$ be $e_i$ if $\pol_i=\altpol$ and $0$ if
$h\notin\bases{\factor{\pol}}$. As the bases are not required to be
irreducible, factorizations are not unique. 

We assume that bases and exponents are non-zero,
$\factor{1}=\{1^1\}$, and $1^k\notin\factor{\pol}$ for
$\pol\neq 1$. For $\factor{\pol}=\{\pol_1^{e_1},\ldots,\pol_n^{e_n}\}$, 
this is expressed by the reduction $\factor{\pol}^{\mathrm{red}}=\{1^1\}$ 
if $n>0$ and $\pol_i=1$ or $e_i=0$ for all $1\leq i \leq n$, and 
$\factor{\pol}^{\mathrm{red}}=\factor{\pol}\setminus\{\pol_i^{e_i}\ |\ \pol_i=1 \vee e_i=0\}$
otherwise.

\paragraph{Operations on factorizations.} Instead of applying arithmetic operations on two polynomials $\pol_1$ and 
$\pol_2$ directly, we operate on their factorizations $\factor{\pol_1}$ and $\factor{\pol_2}$.
We use the following operations on factorizations: $\factor{\pol_1} \fcm \factor{\pol_2}$ factorizes a (not necessarily least) common multiple of $\pol_1$ and $\pol_2$, $\factor{\pol_1}\fcd\factor{\pol_2}$ a (not necessarily greatest) common divisor, whereas the binary operations $\fmult$, $\fdiv$ and $\fplus$ correspond to multiplication, division\footnote{$\factor{\pol_1}\fdiv \factor{\pol_2}$ is a factorization of $\pol_1/\pol_2$ only if $\factor{\pol_1}$ and $\factor{\pol_2}$ are sufficiently refined and $\pol_2$ divides $\pol_1$.} and addition, respectively. Due to space limitations, we omit in the remaining of this paper the trivial cases involving $\factor{0}$. Therefore we define
\[\begin{array}{lcl}
	\factor{\pol_1}\fcm \factor{\pol_2} & =  & \{\altpol^{\max(\exp(\altpol,\factor{\pol_1}), \exp(\altpol,\factor{\pol_2}))}\ |\ \altpol\in \bases{\factor{\pol_1}}\cup \bases{\factor{\pol_2}}\}^{\mathrm{red}}\\
	\factor{\pol_1}\fcd \factor{\pol_2} & =  & \{ \altpol^{\min(\exp(\altpol,\factor{\pol_1}), \exp(\altpol,\factor{\pol_2}))}\ |\ \altpol{=}1 \vee \altpol{\in}\bases{\factor{\pol_1}}{\cap}\bases{\factor{\pol_2}}\}^{\mathrm{red}}\\
	\factor{\pol_1}\fmult \factor{\pol_2} & =  & \{\altpol^{\exp(\altpol,\factor{\pol_1})+\exp(\altpol,\factor{\pol_2})}\ |\ \altpol\in \bases{\factor{\pol_1}}\cup \bases{\factor{\pol_2}}\}^{\mathrm{red}}\\
	\factor{\pol_1}\fdiv \factor{\pol_2} & =  & \{\altpol^{\max(0,e-\exp(\altpol,\factor{\pol_2}))}\ |\ \altpol^{e}\in \factor{\pol_1}\}^{\mathrm{red}}\\
	\factor{\pol_1}\fplus \factor{\pol_2} & =  & D \fmult \bigl\{\bigl(\prod_{\pol_1'\in(\factor{\pol_1}\fdiv D)}\ \pol_1'\bigr) + \bigl(\prod_{\pol_2'\in(\factor{\pol_2}\fdiv D)}\ \pol_2'\bigr)\bigr\}^{\mathrm{red}}\\
%	\factor{\pol_1}\fminus \factor{\pol_2} & =  & D \fmult \{(\Pi_{\pol_1'\in\factor{\pol_1}\fdiv D}\ \pol_1') - (\Pi_{\pol_2'\in\factor{\pol_2}\fdiv D}\ \pol_2')\}^{\mathrm{red}}
\end{array}\]
where $D=\factor{\pol_1}\fcd\factor{\pol_2}$ and $\max(a,b)$ ($\min(a,b)$) equals $a$ if $a\geq b$ ($a\leq b$) and $b$ otherwise. Example~\ref{exa:algo_gcd} illustrates
the application of the above operations.

\paragraph{Operations on rational functions.} We represent a rational function
$\frac{\pol_1}{\pol_2}$ by separate factorizations
$\factor{\pol_1}$ and $\factor{\pol_2}$ for the numerator $\pol_1$ and
the denominator $\pol_2$, respectively. For multiplication
$\frac{\pol_1}{\pol_2}=\frac{\altpol_1}{\altpol_2}\cdot\frac{q_1}{q_2}$,
we compute $\factor{\pol_1}=\factor{\altpol_1}\fmult \factor{q_1}$ and
$\factor{\pol_2}=\factor{\altpol_2}\fmult \factor{q_2}$.  Division is
reduced to multiplication according to
$\frac{\altpol_1}{\altpol_2}\colon\frac{q_1}{q_2} =
\frac{\altpol_1}{\altpol_2}\cdot\frac{q_2}{q_1}$.

For the addition
  $\frac{\pol_1}{\pol_2}=\frac{\altpol_1}{\altpol_2}+\frac{q_1}{q_2}$,
  we compute $\pol_2$ with $\factor{\pol_2}=\factor{\altpol_2}\fcm
  \factor{q_2}$ as a common multiple of $\altpol_2$ and $q_2$, such that $\pol_2=\altpol_2\cdot\altpol_2'$
  with $\factor{\altpol_2'}=\factor{\pol_2}\fdiv\factor{\altpol_2}$, and
$\pol_2=q_2\cdot q_2'$
  with $\factor{q_2'}=\factor{\pol_2}\fdiv\factor{q_2}$. 
For the numerator $\pol_1$ we first determine a common
  divisor $d$ of $\altpol_1$ and $q_1$ by
  $\factor{d}=\factor{\altpol_1}\fcd \factor{q_1}$, such that
  $\altpol_1=d\cdot\altpol_1'$ with $\factor{\altpol_1'}=\factor{\altpol_1}\fdiv
  \factor{d}$, and 
  $q_1=d\cdot q_1'$ with $\factor{q_1'}=\factor{q_1}\fdiv
  \factor{d}$.
 The numerator $\pol_1$ is
  $d\cdot(\altpol_1'\cdot\altpol_2'+q_1'\cdot q_2')$ with factorization $\factor{d}\fmult (\factor{\altpol_1'}\fmult\factor{\altpol_2'}\fplus \factor{q_1'}\fmult\factor{q_2'})$.

\begin{algorithm}[\ifcomment h \else t \fi]
\begin{newalgorithm}{GCD}{factorization $\factor{\pol_1}$, factorization $\factor{\pol_2}$}
  \State $G\colonequals (\factor{\pol_1}\fcd \factor{\pol_2})$\label{algo:gcd:init1}\\
  \State $F_i\colonequals \factor{\pol_i}\fdiv G$ and $F_i'\colonequals \{1^1\}$ for $i=1,2$\label{algo:gcd:init2}\\
  \While{exists $r_1^{e_1}\in F_1$ with $r_1\not=1$}\label{algo:gcd:while1}%
    \State $F_1\colonequals F_1\setminus\{r_1^{e_1}\}$\label{algo:gcd:remove1}\\
    \While{$r_1\not= 1$ and exists $r_2^{e_2}\in F_2$ with $r_2\not=1$}\label{algo:gcd:while2}%
      \State $F_2\colonequals F_2\setminus\{r_2^{e_2}\}$\label{algo:gcd:remove2}\\
      \Ifx{$\neg \mathrm{irreducible}(r_1) \lor \neg \mathrm{irreducible}(r_2)$}{$\pol\colonequals\cgcd(r_1, r_2)$} \label{algo:gcd:common_gcd}
      \Elsex{$\pol\colonequals 1$}
      \If{$\pol = 1$} \label{algo:gcd:case_gcd_one}%
        \State $F_2'\colonequals F_2'\fmult \{r_2^{e_2}\}$\\
      \Else \label{algo:gcd:case_gcd_not_one}%
        \State $r_1\colonequals\frac{r_1}{\pol}$\label{algo:gcd:considered_pairs_update_end1}\\
        \State $F_i\colonequals F_i\fmult \{\pol^{e_i-\min(e_1, e_2)}\}$ for $i=1,2$\label{algo:gcd:considered_pairs_update_end2}\\
        \State $F_2'\colonequals F_2'\fmult \{(\frac{r_2}{\pol})^{e_2}\}$\\
        \State $G\colonequals G\fmult \{\pol^{\min(e_1,e_2)}\}$\\
      \EndIf\label{algo:gcd:end_refinement_step}%
    \EndWhile%
    \State $F_1'\colonequals F_1' \fmult \{r_1^{e_1}\}$\label{algo:gcd:add_to_pol_1_prime}\\
    \State $F_2\colonequals F_2\fmult F_2'$\\    
    \State $F_2'\colonequals \{1^1\}$\label{algo:gcd:reset_pol_2}\\
  \EndWhile%
  \Return{($F_1', F_2, G$)}
\end{newalgorithm}
\caption{gcd computation with factorization refinement}
\label{lst:gcd}
\end{algorithm}

The rational function $\frac{\pol_1}{\pol_2}$ resulting from the 
addition is further simplified by cancellation, \ie dividing
$\pol_1$ and $\pol_2$ by their greatest common divisor ($\cgcd$) $g$. 
Given the factorizations $\factor{\pol_1}$ and $\factor{\pol_2}$, 
Algorithm~\ref{lst:gcd} calculates the factorizations $\factor{g}$,
$\factor{\frac{\pol_1}{g}}$, and $\factor{\frac{\pol_2}{g}}$.

Intuitively, the algorithm maintains the fact that $G\fmult F_1\fmult
F_1'$ is a factorization of $\pol_1$, where $G$ contains common
factors of $\pol_1$ and $\pol_2$, $F_1$ is going to be checked whether
it contains further common factors, and $F_1'$ does not contain any common
factors.  In the outer while-loop, an element $r_1^{e_1}$ to be
checked is taken from $F_1$. In the inner while-loop, a factorization
$G\fmult F_2\fmult F_2'$ of $\pol_2$ is maintained such that $F_2'$
does not contain any common factors with $r_1$, and $F_2$ is still to
be checked.

Now we explain the algorithm in more detail. Initially, a
factorization $G$ of a common divisor of $\pol_1$ and $\pol_2$ is set
to $\factor{\pol_1}\fcd \factor{\pol_2}$\linex{algo:gcd:init1}. The
remaining factors of $\pol_1$ and $\pol_2$ are stored in $F_1$ resp.
$F_2$. The sets $F_1'$ and $F_2'$ contain factors of $\pol_1$ resp.
$\pol_2$ whose greatest common divisor is $1$\linex{algo:gcd:init2}.
The algorithm now iteratively adds further common divisors of $\pol_1$
and $\pol_2$ to $G$ until it is a factorization of their $\cgcd$. For this
purpose, we consider for each factor in $F_1$ all factors in $F_2$ and
calculate the $\cgcd$ of their bases using standard $\cgcd$
computation for polynomials\linex{algo:gcd:common_gcd}. Note that the main concern of
Algorithm~\ref{lst:gcd} is to avoid the application of this expensive operation as far as possible and to apply it to preferably simple polynomials otherwise. Where the latter is
entailed by the idea of using factorizations, the former can be
achieved by excluding pairs of factors for which we can cheaply decide
that both are \emph{irreducible}, \ie they have no non-trivial
divisors. If factors $r_1^{e_1}\in F_1$ and $r_2^{e_2}\in F_2$ with
$\pol\colonequals \cgcd(r_1, r_2)=1$ are found, we just
shift $r_2^{e_2}$ from $F_2$ to $F_2'$\linex{algo:gcd:case_gcd_one}.
Otherwise, we can add $\pol^{\min(e_1,e_2)}$, which is the $\cgcd$ of
$r_1^{e_1}$ and $r_2^{e_2}$, to $G$ and extend the factors $F_1$ resp.
$F_2$, which could still contain common divisors, by
$\pol^{e_1-\min(e_1, e_2)}$ resp. $\pol^{e_2-\min(e_1,
  e_2)}$\linex{algo:gcd:case_gcd_not_one}. Furthermore, $F_2'$ obtains
the new factor $(\frac{r_2}{\pol})^{e_2}$, which has certainly no
common divisor with any factor in $F_1'$. Finally, we set the basis
$r_1$ to $\frac{r_1}{\pol}$, excluding the just found common divisor.
If all factors in $F_2$ have been considered for common divisors with
$r_1$, we can add it to $F_1'$ and continue with the next factor in
$F_1$, for which we must reconsider all factors in $F_2'$ and,
therefore, shift them to
$F_2$\linexx{algo:gcd:add_to_pol_1_prime}{algo:gcd:reset_pol_2}. The
algorithm terminates, if the last factor of $F_1$ has been processed,
returning the factorizations $\factor{g}$,
$\factor{\frac{\pol_1}{g}}$ and $\factor{\frac{\pol_2}{g}}$, which
we can use to refine the factorizations of $\pol_1$ and $\pol_2$ via
$\factor{\pol_1} \colonequals \factor{\frac{\pol_1}{g}} \fmult G$ and
$\factor{\pol_2} \colonequals \factor{\frac{\pol_2}{g}} \fmult G$.

\begin{example}
\label{exa:algo_gcd}
Assume we want to apply Algorithm~\ref{lst:gcd} to the factorizations $\factor{xyz}=\{(xyz)^1\}$ and $\factor{xy}=\{(x)^1,\ (y)^1\}$. We initialize $G=F_1'=F_2'=\{(1)^1\}$, $F_1=\factor{xyz}$ and $F_2=\factor{xy}$. First, we choose the factors $(r_1)^{e_1}=(xyz)^1$ and $(x)^1$ and remove them from $F_1$ resp. $F_2$. The $\cgcd$ of their bases is $x$, hence we only update $r_1$ to $(yz)^1$ and $G$ to $\{(x)^1\}$. Then we remove the next and last element $(y)^1$ from $F_2$. Its basis and $r_1$ have the $\cgcd$ $y$ and we therefore update
$r_1$ to $(z)^1$ and $G$ to $\{(x)^1,\ (y)^1\}$. Finally, we add $(z)^1$ to $F_1'$ and return the expected result $(\{(z)^1\},\ \{(1)^1\},\ \{(x)^1,\ (y)^1\})$. Using these results,
we can also refine $\factor{xyz}=F_1'\fmult G=\{(x)^1,\ (y)^1,\ (z)^1\}$ and $\factor{xy}=F_2\fmult G=\{(x)^1,\ (y)^1\}$.
\end{example}

\newcommand{\algogcdtheorem}{
  Let $p_1$ and $p_2$ be two polynomials with factorizations $\factor{p_1}$ resp. $\factor{p_2}$. Applying Algorithm~\ref{lst:gcd} to these factorizations results in $\gcd(\factor{p_1},\, \factor{p_2}) = (\factor{r_1},\, \factor{r_2},\, G)$ with $G$ being a factorization of the greatest common divisor $g$ of $p_1$ and $p_2$, and $\factor{r_1}$ and $\factor{r_2}$ being factorizations of $\frac{p_1}{g}$ resp. $\frac{p_2}{g}$.
}

\begin{theorem}\label{theo:algogcd}
  \algogcdtheorem
\end{theorem}

\noindent The proof of this theorem can be found in the appendix.

  \section{Experiments}
\label{sec:experiments}

We developed a \cpp prototype implementation of our approach using the arithmetic library \ginac~\cite{GiNaC02}. The prototype is available on the project homepage\footnote{\homepage}. Moreover, we implemented the state-elimination approach used by PARAM~\cite{PARAM10} using our optimized factorization approach to provide a more distinct comparison.
All experiments were run on an Intel Core 2 Quad CPU 2.66 GHz with 4 GB of memory. We defined a timeout ($\TO$) of 14 hours (50400 seconds) and a memory bound ($\MO$) of 4 GB.
\ifshort
We report on three case studies; a more distinct description and the specific instances we used are available at our homepage.

The \emph{bounded retransmission protocol} (BRP)~\cite{HSV94} models the sending of files via an unreliable network, manifested in two lossy channels for sending and acknowledging the reception. This model is parametrized in the probability of reliability of those channels.
The \emph{crowds protocol} (CROWDS)~\cite{RR98} is designed for anonymous network communication using random routing, parametrized in how many members are ``good'' or ``bad'' and the probability if a good member delivers a message or randomly routes it to another member.
\emph{NAND multiplexing} (NAND)~\cite{HJ02} models how reliable computations are obtained using unreliable hardware by having a certain number of copies of a NAND unit all doing the same job. Parameters are the probabilities of faultiness of the units and of erroneous inputs.
The experimental setting includes our SCC-based approach as described in Section~\ref{sec:scc_reduction} using the optimized factorization of polynomials as in Section~\ref{sec:factorization} (SCC MC), the state elimination as in PARAM but also using the approach of Section~\ref{sec:factorization} (STATE ELIM) and the PARAM tool itself.\footnote{Note that no bisimulation reduction was applied to any of the input models, which would improve the feasibility of all approaches likewise.} 
For all instances we list the number of states and transitions; for each tool we give the running time in seconds and the memory consumption in MB; the best time is \textbf{boldfaced}. Moreover, for our approaches we list the number of polynomials which are intermediately stored. 
%The best running time for each instance is marked boldface. 

\begin{table}[h]
  \centering
  \label{tab:results}
%  \caption{Experimental results}
  \small
  \setlength{\tabcolsep}{3pt}
  \scalebox{0.83}{\footnotesize
  \begin{tabular}{lrrrrrrrrrrr}
    \toprule
            &\multicolumn{2}{c}{Graph} &\multicolumn{3}{c}{SCC MC} &\multicolumn{3}{c}{STATE ELIM} &\multicolumn{2}{c}{PARAM}\\
    \cmidrule(lr){2-3}\cmidrule(lr){4-6}\cmidrule(lr){7-9}\cmidrule(lr){10-11}
    Model & States & Trans. & Time & Poly & Mem &Time & Poly &Mem &Time &Mem \\
    \midrule
    %BRP &679 &867 &0.05 &591 &6.00 &0.10 &1607 &7.81 &0.48 &4.57\\
    %BRP &888 &1155 &0.16 &835 &6.96 &0.16 &1987 &8.81 &1.21 &5.52\\
    %BRP &1097 &1443 &0.69 &1079 &9.80 &0.25 &2320 &9.53 &2.24 &6.67\\
    %BRP &1306 &1731 &2.62 &1323 &19.14 &0.32 &2653 &11.47 &3.67 &8.24\\
    %BRP &1351 &1731 &0.26 &1167 &8.09 &0.39 &3287 &15.04 &3.97 &7.09\\
    %BRP &1768 &2307 &1.92 &1651 &13.43 &0.76 &4051 &19.75 &10.66 &10.91\\
%    BRP &2185 &2883 &15.42 &2135 &45.31 &1.10 &4720 &23.51 &20.03 &16.26\\
%    BRP &2602 &3459 &118.75 &2619 &272.11 &1.69 &5389 &24.50 &32.98 &21.52\\
%    BRP &2695 &3459 &1.86 &2319 &16.04 &1.97 &6647 &40.04 &36.49 &17.38\\
    BRP &3528 &4611 &29.05 &3283 &48.10 &\textbf{4.33} &8179 &61.17 &98.99 &32.90\\
    BRP &4361 &5763 &511.50 &4247 &501.71 &\textbf{6.87} &9520 &78.49 &191.52 &58.43\\
%    BRP &5194 &6915 &--- &--- &\MO &8.64 &10861 &80.31 &318.11 &85.00\\
%    BRP &5383 &6915 &14.81 &4623 &47.00 &12.19 &13367 &161.80 &356.04 &63.21\\
    BRP &7048 &9219 &548.73 &6547 &281.86 &\textbf{25.05} &16435 &216.05 &988.28 &142.66\\
%    BRP &8713 &11523 &--- &--- &\MO &49.90 &19120 &276.57 &1856.63 &279.73\\
%    BRP &10378 &13827 &--- &--- &\MO &67.19 &21805 &392.08 &3137.53 &431.35\\
    BRP &10759 &13827 &147.31 &9231 &176.89 &\textbf{85.54} &26807 &682.24 &3511.96 &304.07\\
%    BRP &14088 &18435 &10399.50 &13070 &2066.94 &209.77 &32947 &1018.75 &9623.14 &770.46\\
%    BRP &17417 &23043 &--- &--- &\MO &382.62 &38320 &1482.87 &18272.50 &1614.86\\
 %   BRP &20746 &27651 &--- &--- &\MO &\textbf{580.72} &43693 &1648.77 &30273.20 &2594.04\\
    BRP &21511 &27651 &1602.53 &18443 &776.48 &\textbf{718.66} &53687 &3134.59 &34322.60 &1757.12\\
    %CROWDS &398 &576 &0.05 &442 &7.14 &0.03 &315 &7.14 &0.56 &5.86\\
    %CROWDS &1774 &2612 &0.18 &1414 &4.95 &0.18 &934 &7.45 &9.02 &5.88\\
%    CROWDS &5205 &7723 &0.43 &3355 &9.51 &0.70 &2152 &9.51 &75.62 &9.36\\
%    CROWDS &17394 &25972 &0.92 &2862 &16.64 &3.23 &5289 &16.62 &836.69 &22.78\\
%    CROWDS &43788 &65626 &2.47 &5619 &35.39 &10.50 &10480 &32.75 &6019.88 &63.01\\
%    CROWDS &73239 &109957 &4.16 &8361 &52.91 &20.53 &15386 &50.44 &18197.60 &105.73\\
%    CROWDS &115485 &173623 &6.84 &11777 &95.93 &37.87 &21662 &75.64 &\TO &---\\
%    CROWDS &210059 &316317 &13.24 &18661 &189.78 &83.23 &34255 &133.31 &\TO &---\\
%begin new
    CROWDS &198201 &348349 &\textbf{60.90} &13483 &140.15 &243.07 &27340 &133.91 &46380.00 &227.66\\
    CROWDS &482979 &728677 &\textbf{35.06} &35916 &478.85 &247.75 &65966 &297.40 &\TO &--- \\
    CROWDS &726379 &1283297 &\textbf{223.24} &36649 &515.61 &1632.63 &73704 &477.10 &\TO &---\\
    CROWDS &961499 &1452537 &\textbf{81.88} &61299 &1027.78 &646.76 &112452 &589.21 &\TO &--- \\
    CROWDS &1729494 &2615272 &\textbf{172.59} &97655 &2372.35 &1515.63 &178885 &1063.15 &\TO &--- \\
%    CROWDS &2885214 &4366132 &--- &--- &\MO &2956.16 &267816 &1785.84 &\TO &--- \\
%end new
    %CROWDS &1200 &2038 &0.53 &1519 &7.22 &0.22 &720 &7.22 &5.34 &5.87\\
%    CROWDS &8655 &14953 &4.15 &8747 &13.21 &\textbf{3.24} &2943 &11.96 &139.70 &10.44\\
%    CROWDS &37293 &65011 &16.69 &33549 &40.23 &21.72 &8148 &30.61 &1977.95 &35.39\\
%    CROWDS &1507767 &2670565 &\textbf{416.12} &65534 &1127.40 &4878.02 &133423 &1025.21 &\TO &---\\
    CROWDS &2888763 &5127151 &\textbf{852.76} &110078 &2345.06 &12326.80 &224747 &2123.96 &\TO &---\\
%    CROWDS &6565 &15143 &43.23 &2646 &11.33 &5.61 &1743 &11.11 &135.72 &8.55\\
%    CROWDS &111296 &26144 &780.80 &12336 &91.82 &224.60 &7186 &88.11 &9515.02 &98.70\\
%    CROWDS &990603 &2351961 &7472.86 &65209 &815.34 &4847.21 &20007 &737.74 &\TO \ (24h) &---\\
%    CROWDS &19230 &55948 &511.86 &6352 &25.36 &37.68 &2826 &21.24 &1019.42 &19.32\\
%    CROWDS &592062 &1754860 &\TO &--- &- &2902.03 &12474 &479.08 &\TO &--- &\\
    NAND &7393 &11207 &8.35 &15688 &114.60 &17.02 &140057 &255.13 &\textbf{5.00} &10.67\\
    NAND &14323 &21567 &39.71 &25504 &366.79 &59.60 &405069 &926.33 &\textbf{15.26} &16.89\\
    NAND &21253 &31927 &100.32 &35151 &795.31 &121.40 &665584 &2050.67 &\textbf{29.51} &24.45\\
    NAND &28183 &42287 &208.41 &44799 &1405.16 &218.85 &925324 &3708.27 &\textbf{50.45} &30.47\\
%    NAND &35113 &52647 &352.09 &54445 &2047.66 &364.09 &1184848 &3696.39 &\textbf{78.19} &40.51\\
    NAND &78334 &121512 &\textbf{639.29} &184799 &3785.11 &--- &--- &\MO &1138.82 &111.58\\
    \bottomrule
  \end{tabular}}
\end{table}

For BRP, STATE ELIM always outperforms PARAM and SCC MC by up to two orders of 
magnitude. On larger instances, SCC MC is faster than PARAM while on smaller 
ones PARAM is faster and has a smaller memory consumption.

In contrast, the crowds protocol always induces a nested SCC structure, which is 
very hard for PARAM since many divisions of polynomials have to be carried out. 
On larger benchmarks, it is therefore outperformed by more than three orders of 
magnitude while SCC MC performs best. Please note that this is measured by the 
timeout. In fact, we were not able to retrieve results for PARAM on the larger 
crowds instances.

To give an example where PARAM performs mostly better than our approaches, we 
consider NAND. Its graph consists of single paths, inducing a high number of 
polynomials we store. Our implementation offers the possibility to limit the number
of stored polynomials, which decreases the memory consumption at the price 
of losing information about the factorizations. However, an efficient strategy 
to manage this bounded pool of polynomials is not yet implemented. Therefore, we 
refrain from presenting experimental results for this scenario. 

\fi

  \section{Conclusion and Future Work}
\label{sec:conclusion}
We presented a new approach to verify parametric Markov chains together with an improved factorization of polynomials. We were able to highly
improve the scalability in comparison to existing approaches. Future work will
be dedicated to the actual parameter synthesis. First, we want to incorporate
interval constraint propagation~\cite{FHT+:2007} in order to
provide reasonable intervals for the parameters where properties are satisfied
or violated. Moreover, we are going to investigate the possibility of extending our approaches to models with costs.

  \bibliographystyle{splncs}
  \bibliography{literature}

  \newpage
\section*{Appendix}

\begin{mytheorem}{\ref{theo:abstraction}}
  \pdtmcabstractiontheorem
\end{mytheorem}
\begin{proof}\label{app:proof_pdtmc_abstraction}
  First note that all initial states and absorbing states in $\PDTMCM$ are also states of the
  abstraction.

As the bottom SCCs are the absorbing states in $T$,
the probability of reaching a state in $T$ is $1$. The probability $p\abs^{\PDTMCM}(\sinit, \sinit)$
can therefore be expressed w.\,r.\,t.\ the probabilities of reaching
an absorbing state without revisiting $\sinit$:
\begin{align}\label{eq:selfloop}
  p\abs^{\PDTMCM}(\sinit,\sinit)=1-\sum_{\starget\in T} p\abs^{\PDTMCM}(\sinit,\starget).
\end{align}
\noindent To reduce notation, we define the set of paths $\Rloop$ looping on $\sinit$ and 
the set of paths $\Rout$ going to some $\starget\in T$ without revisiting $\sinit$.
\begin{align}
  \Rloop = {}&\{\sinit \seriesPath{s}{n} \sinit \in \pathsfin^{\PDTMCM}\ |\
  s_i\notin \{\sinit\}\cup T, 1 \leq i \leq n\}\label{eq:app_rloop}\\
  \Rout = {}&\{\sinit \seriesPath{s}{n} \starget \in \pathsfin^{\PDTMCM}\ |\
  s_i\notin \{\sinit\}\cup T, 1 \leq i \leq n, \starget\in T\}\label{eq:app_rout}
\end{align}
As the self-loop on $\sinit$ represents the paths of $\Rloop$, it holds that
\begin{align}\label{eq:pabsPaths}
  p\abs^{\PDTMCM}(\sinit, \sinit)=\pr(\Rloop).
\end{align}
We now have:
{\allowdisplaybreaks
\begin{alignat*}{2}
  &&&\pr^{\PDTMCM}\bigl(\pathsfin^{\PDTMCM}(\sinit, \starget)\bigr)\\
  &{=}\quad&&\pr^{\PDTMCM}\Bigl(\bigcup_{i=0}^\infty \{\pi_1\cdot\cdots\cdot\pi_i\cdot\pi_{\mathrm{out}} \ |\ \pi_j\in\Rloop,1\leq j\leq i;\ \pi_{\mathrm{out}}\in\Rout\}\Bigr)\\
  &{=}&&\sum_{i=0}^\infty\pr^{\PDTMCM}\bigl(\{\pi_1\cdot\cdots\cdot\pi_i\cdot\pi_{\mathrm{out}}\ |\ \pi_j\in\Rloop,1\leq j\leq i;\ \pi_{\mathrm{out}}\in\Rout\}\bigr)\\
  &{=}&&\sum_{i=0}^\infty\bigl(\pr^{\PDTMCM}(\Rloop)\bigr)^i\cdot\pr^{\PDTMCM}(\Rout)\\
  &{=}&&\sum_{i=0}^\infty \bigl(p\abs^{\PDTMCM}(\sinit,\sinit)\bigr)^i \cdot\pr^{\PDTMCM}(\Rout)\quad\text{ (Equation~\eqref{eq:pabsPaths})}\\
  &{=}&&\frac{1}{1-p\abs^{\PDTMCM}(\sinit,\sinit)}\cdot\pr^{\PDTMCM}(\Rout)\quad\text{ (Geometric Series)}\\
  &{=}&&\frac{1}{\sum\limits_{s_\mathrm{out}\in T}p\abs^{\PDTMCM}(\sinit,s_\mathrm{out})}\cdot\pr^{\PDTMCM}(\Rout)\quad\text{ (Equation~\eqref{eq:selfloop})}\\
  &{=}&&\frac{1}{\sum\limits_{s_\mathrm{out}\in T}p\abs^{\PDTMCM}(\sinit,s_\mathrm{out})}\cdotp p\abs^{\PDTMCM}(\sinit,\starget)\quad\text{ (Definition~\ref{def:abstractPMC})}\\
  &{=}&&P\abs(\sinit,\starget)\quad\text{ (Definition~\ref{def:abstractPMC})}\\
  &{=}&&\pr^{\PDTMCM\abs}\bigl(\pathsfin^{\PDTMCM\abs}(\sinit,\starget)\bigr)\ .
\end{alignat*}}
As the probabilities of reaching the absorbing states from initial states coincide in $\PDTMCM$ and $\PDTMCM\abs$, our abstraction is valid.
\end{proof}

\begin{mytheorem}{\ref{theo:algogcd}}
  \algogcdtheorem
\end{mytheorem}
\begin{proof}\label{app:proof_gcd_correctness}
  We denote the product of a factorization $\factor{p}$ by $\product{\factor{p}} =\prod_{q^e \in \factor{p}}q^e$ and the standard greatest common divisor computation for polynomials 
by $\cgcd$.

We define the following Hoare-style assertion network:

\begin{newalgorithm}{GCD}{factorization $\factor{\pol_1}$, factorization $\factor{\pol_2}$}
\State \hspace*{-3.5ex}{\scriptsize$ \{\textit{true}\}$}\\
  \State $G\colonequals (\factor{\pol_1}\fcd \factor{\pol_2})$\label{algo:gcd:init1}\\
\State \hspace*{-3.5ex}  {\scriptsize$ \{G = \factor{\pol_1}\fcd \factor{\pol_2}\}$}\\
  \State $F_i\colonequals \factor{\pol_i}\fdiv G$ and $F_i'\colonequals \{1^1\}$ for $i=1,2$\label{algo:gcd:init2}\\
\State \hspace*{-3.5ex}  {\scriptsize$ \{\factor{\pol_1}=G\fmult F_1\fmult F_1'\wedge \factor{\pol_2}=G\fmult F_2\fmult F_2'\wedge \product{F_1'}=1\wedge\product{F_2'}=1\}$}\label{algo:while_pre}\\
  \While{exists $r_1^{e_1}\in F_1$ with $r_1\not=1$}\label{algo:gcd:while1}%
\State \hspace*{-7ex}  {\scriptsize$ \{\factor{\pol_1}=G\fmult F_1\fmult F_1'\wedge \factor{\pol_2}=G\fmult F_2\fmult F_2'\wedge \newline \phantom{\{}\cgcd(\product{F_1'},\product{F_2\fmult F_2'})=1\wedge\cgcd(r_1^{e_1},\product{F_2'})=1\wedge r_1^{e_1}\in F_1\}$}\label{algo:while_pre1}\\
    \State $F_1\colonequals F_1\setminus\{r_1^{e_1}\}$\label{algo:gcd:remove1}\\
\State \hspace*{-7ex}  {\scriptsize$ \{\factor{\pol_1}=G\fmult F_1\fmult F_1'\fmult\{r_1^{e_1}\}\wedge \factor{\pol_2}=G\fmult F_2\fmult F_2'\wedge \newline\phantom{\{}\cgcd(\product{F_1'},\product{F_2\fmult F_2'})=1\wedge\cgcd(r_1^{e_1},\product{F_2'})=1\}$}\\
    \While{$r_1\not= 1$ and exists $r_2^{e_2}\in F_2$ with $r_2\not=1$}\label{algo:gcd:while2}%
\State \hspace*{-10.5ex}  {\scriptsize$ \{\factor{\pol_1}=G\fmult F_1\fmult F_1'\fmult\{r_1^{e_1}\}\wedge \factor{\pol_2}=G\fmult F_2\fmult F_2'\wedge \newline\phantom{\{}\cgcd(\product{F_1'},\product{F_2\fmult F_2'})=1\wedge \cgcd(r_1^{e_1},\product{F_2'})=1\wedge r_2^{e_2}\in F_2\}$}\\
      \State $F_2\colonequals F_2\setminus\{r_2^{e_2}\}$\label{algo:gcd:remove2}\\
\State \hspace*{-10.5ex}  {\scriptsize$ \{\factor{\pol_1}=G\fmult F_1\fmult F_1'\fmult\{r_1^{e_1}\}\wedge \factor{\pol_2}=G\fmult F_2\fmult F_2'\fmult\{r_2^{e_2}\}\wedge\newline\phantom{\{}\cgcd(\product{F_1'},\product{F_2\fmult F_2'\fmult\{r_2^{e_2}\}})=1\wedge \cgcd(r_1^{e_1},\product{F_2'})=1\}$}\\
      \Ifx{$\neg \mathrm{irreducible}(r_1) \lor \neg \mathrm{irreducible}(r_2)$}{$\pol\colonequals\ \cgcd(r_1, r_2)$} \label{algo:gcd:common_gcd}
      \Elsex{$\pol\colonequals 1$} \label{algo:gcd:common_gcd2}
\State \hspace*{-10.5ex}  {\scriptsize$ \{\factor{\pol_1}=G\fmult F_1\fmult F_1'\fmult\{r_1^{e_1}\}\wedge \factor{\pol_2}=G\fmult F_2\fmult F_2'\fmult\{r_2^{e_2}\}\wedge \newline \phantom{\{}\cgcd(\product{F_1'},\product{F_2\fmult F_2'\fmult\{r_2^{e_2}\}})=1\wedge \cgcd(r_1^{e_1},\product{F_2'})=1\wedge g=\cgcd(r_1,r_2)\}$}\label{algo:ite_pre}\\
      \If{$\pol = 1$} \label{algo:gcd:case_gcd_one}%
\State \hspace*{-14ex}  {\scriptsize$ \{\factor{\pol_1}=G\fmult F_1\fmult F_1'\fmult\{r_1^{e_1}\}\wedge \factor{\pol_2}=G\fmult F_2\fmult F_2'\fmult\{r_2^{e_2}\}\wedge \newline \phantom{\{}\cgcd(\product{F_1'},\product{F_2\fmult F_2'\fmult\{r_2^{e_2}\}})=1\wedge \cgcd(r_1^{e_1},\product{F_2'})=1\wedge \cgcd(r_1,r_2)=1\}$}\label{algo:ite_pre1}\\
        \State $F_2'\colonequals F_2'\fmult \{r_2^{e_2}\}$\\
\State \hspace*{-14ex}  {\scriptsize$ \{\factor{\pol_1}=G\fmult F_1\fmult F_1'\fmult\{r_1^{e_1}\}\wedge \factor{\pol_2}=G\fmult F_2\fmult F_2'\wedge \newline\phantom{\{}\cgcd(\product{F_1'},\product{F_2\fmult F_2'})=1\wedge \cgcd(r_1^{e_1},\product{F_2'})=1\}$}\label{algo:ite_post1}\\
      \Else \label{algo:gcd:case_gcd_not_one}%
\State \hspace*{-14ex}  {\scriptsize$ \{\factor{\pol_1}=G\fmult F_1\fmult F_1'\fmult\{r_1^{e_1}\}\wedge \factor{\pol_2}=G\fmult F_2\fmult F_2'\fmult\{r_2^{e_2}\}\wedge \newline \phantom{\{}\cgcd(\product{F_1'},\product{F_2\fmult F_2'\fmult\{r_2^{e_2}\}})=1\wedge \cgcd(r_1^{e_1},\product{F_2'})=1\wedge g=\cgcd(r_1,r_2)\}$}\label{algo:ite_pre2}\\
        \State $r_1\colonequals\frac{r_1}{\pol}$\label{algo:gcd:considered_pairs_update_end1}\\
\State \hspace*{-14ex}  {\scriptsize$ \{\factor{\pol_1}=G\fmult F_1\fmult F_1'\fmult\{(r_1\cdot g)^{e_1}\}\wedge \factor{\pol_2}=G\fmult F_2\fmult F_2'\fmult\{r_2^{e_2}\}\wedge \newline \phantom{\{}\cgcd(\product{F_1'},\product{F_2\fmult F_2'\fmult\{r_2^{e_2}\}})=1\wedge \cgcd((r_1\cdot g)^{e_1},\product{F_2'})=1\wedge g=\cgcd((r_1\cdot g),r_2)\}$}\\
        \State $F_i\colonequals F_i\fmult \{\pol^{e_i-\min(e_1, e_2)}\}$ for $i=1,2$\label{algo:gcd:considered_pairs_update_end2}\\
\State \hspace*{-14ex}  {\scriptsize$ \{\factor{\pol_1}=G\fmult F_1\fmult F_1'\fmult\{r_1^{e_1}, g^{\min(e_1,e_2)}\}\wedge \factor{\pol_2}=G\fmult F_2\fmult F_2'\fmult\{(\frac{r_2}{g})^{e_2},g^{\min(e_1,e_2)}\}\wedge \newline \phantom{\{}\cgcd(\product{F_1'},\product{F_2\fmult F_2'\fmult\{(\frac{r_2}{g})^{e_2},g^{\min(e_1,e_2)}\}})=1\wedge \cgcd((r_1\cdot g)^{e_1},\product{F_2'})=1\wedge \newline\phantom{\{} g=\cgcd((r_1\cdot g),r_2)\}$}\\
        \State $F_2'\colonequals F_2'\fmult \{(\frac{r_2}{\pol})^{e_2}\}$\\
\State \hspace*{-14ex}  {\scriptsize$ \{\factor{\pol_1}=G\fmult F_1\fmult F_1'\fmult\{r_1^{e_1}, g^{\min(e_1,e_2)}\}\wedge \factor{\pol_2}=G\fmult F_2\fmult F_2'\fmult\{g^{\min(e_1,e_2)}\}\wedge\newline \phantom{\{}\cgcd(\product{F_1'},\product{F_2\fmult F_2'\fmult\{g^{\min(e_1,e_2)}\}})=1\wedge \cgcd((r_1\cdot g)^{e_1},\product{F_2'})=1\}$}\\
        \State $G\colonequals G\fmult \{\pol^{\min(e_1,e_2)}\}$\\
\State \hspace*{-14ex}  {\scriptsize$ \{\factor{\pol_1}=G\fmult F_1\fmult F_1'\fmult\{r_1^{e_1}\}\wedge \factor{\pol_2}=G\fmult F_2\fmult F_2'\wedge \newline \phantom{\{}\cgcd(\product{F_1'},\product{F_2\fmult F_2'})=1\wedge \cgcd(r_1^{e_1},\product{F_2'})=1\}$}\label{algo:ite_post2}\\
      \EndIf\label{algo:gcd:end_refinement_step}%
\State \hspace*{-10.5ex}  {\scriptsize$ \{\factor{\pol_1}=G\fmult F_1\fmult F_1'\fmult\{r_1^{e_1}\}\wedge \factor{\pol_2}=G\fmult F_2\fmult F_2'\wedge \newline \phantom{\{}\cgcd(\product{F_1'},\product{F_2\fmult F_2'})=1\wedge \cgcd(r_1^{e_1},\product{F_2'})=1\}$}\label{algo:ite_post}\\
    \EndWhile%
\State \hspace*{-7ex}  {\scriptsize$ \{\factor{\pol_1}=G\fmult F_1\fmult F_1'\fmult\{r_1^{e_1}\}\wedge \factor{\pol_2}=G\fmult F_2\fmult F_2'\wedge \newline \phantom{\{}\cgcd(\product{F_1'},\product{F_2\fmult F_2'})=1\wedge \cgcd(r_1^{e_1},\product{F_2'})=1\wedge (r_1=1\vee\product{F_2}=1)$}\\
    \State $F_1'\colonequals F_1' \fmult \{r_1^{e_1}\}$\label{algo:gcd:add_to_pol_1_prime}\\
\State \hspace*{-7ex}  {\scriptsize$ \{\factor{\pol_1}=G\fmult F_1\fmult F_1'\wedge \factor{\pol_2}=G\fmult F_2\fmult F_2'\wedge \cgcd(\product{F_1'},\product{F_2\fmult F_2'})=1\}$}\\
    \State $F_2\colonequals F_2\fmult F_2'$\\    
\State \hspace*{-7ex}  {\scriptsize$ \{\factor{\pol_1}=G\fmult F_1\fmult F_1'\wedge \factor{\pol_2}=G\fmult F_2\wedge \cgcd(\product{F_1'},\product{F_2})=1\}$}\\
    \State $F_2'\colonequals \{1^1\}$\label{algo:gcd:reset_pol_2}\\
\State \hspace*{-7ex}  {\scriptsize$ \{\factor{\pol_1}=G\fmult F_1\fmult F_1'\wedge \factor{\pol_2}=G\fmult F_2\wedge \cgcd(\product{F_1'},\product{F_2})=1\wedge \product{F_2'}=1\}$}\label{algo:while_post1}\\
  \EndWhile\label{algo:endwhile}%
\State \hspace*{-3.5ex}  {\scriptsize$ \{\factor{\pol_1}=G\fmult F_1'\wedge \factor{\pol_2}=G\fmult F_2\wedge \cgcd(\product{F_1'},\product{F_2})=1\}$}\label{algo:while_post}\\
  \Return{($F_1', F_2, G$)}
\end{newalgorithm}

The above assertion network is inductive.  
\begin{itemize}
\item For the assignments, their preconditions imply their postconditions
after substituting the assigned expression for the assigned variables.
(For simplicity, we handle the first if-then-else statement in lines
\eqref{algo:gcd:common_gcd}-\eqref{algo:gcd:common_gcd2} also as
atomic assignment.)
\item 
For the if-then-else statement in lines
\eqref{algo:gcd:case_gcd_one}-\eqref{algo:gcd:end_refinement_step},
its precondition \eqref{algo:ite_pre} implies the precondition
\eqref{algo:ite_pre1} of the if-branch if the branching condition
holds, and the precondition \eqref{algo:ite_pre2} of the else-branch
if the condition does not hold. The postconditions
\eqref{algo:ite_post1} and \eqref{algo:ite_post2} of both branches
imply the postcondition \eqref{algo:ite_post} of the if-then-else
statement.
\item 
For the outer while-loop
\eqref{algo:gcd:while1}-\eqref{algo:endwhile}, its precondition
\eqref{algo:while_pre} as well as the postcondition
\eqref{algo:while_post1} of its body imply the precondition
\eqref{algo:while_pre1} of the body if the loop condition holds, and
they both imply the postcondition \eqref{algo:while_post} of the while
loop if the loop condition does not hold.
\item
The inner while loop's inductivity can be shown similarly.
\end{itemize}
That means, the assertion \eqref{algo:while_post} always holds before
returning, implying the correctness of the algorithm.

The algorithm is also complete, since it always terminates:
We can use as ranking function the sum of the degrees of all
polynomials in $F_1$ for the outer loop and in $F_2$ for the inner
loop to show their termination.

\end{proof}

\end{document}